\documentclass[prd,aps,twocolumn,a4paper,showkeys,nofootinbib]{revtex4-2}

\usepackage{amsmath}
\usepackage{amsfonts}
\usepackage{amssymb}	
\usepackage{graphicx}
\usepackage{amsbsy}
\usepackage{color}
\usepackage{commath}
\allowdisplaybreaks
\usepackage{hyperref}
\usepackage{makecell}
\usepackage{multirow}
\usepackage{bm}
\usepackage{orcidlink}

\usepackage[normalem]{ulem}

\def\GMc2{G M_{\odot} c^{-2}}

\def\TEOBResumS{\texttt{TEOBResumS}}

\newcommand\fnp[1]{{\hat{f}_{\varphi #1}^{\rm N _{nc}}}}
\def\NP22{{\fnp{,22}}}

\usepackage{color}
\definecolor{cyan}{rgb}{0,0.9,0.9}
\definecolor{orange}{rgb}{0.9,0.5,0}
\definecolor{magenta}{rgb}{1,0,1}
\definecolor{purple}{rgb}{0.8,0.4,0.8}
\definecolor{gray}{rgb}{0.8242,0.8242,0.8242}
\definecolor{dodgerblue}{rgb}{0.12, 0.56, 1.0}
\definecolor{darkgrey}{rgb}{0.5,0.5,0.5}
\definecolor{darkgreen}{rgb}{0,0.65,0}

\definecolor{colortab1}{rgb}{0.1, 0.1, 1.0}
\definecolor{colortab2}{rgb}{0.9,0,0.1}

\begin{document}
\title{Gravitational spin-orbit coupling through the third-subleading post-Newtonian order: exploring spin-gauge flexibility}

\author{Andrea \surname{Placidi}\orcidlink{0000-0001-8032-4416}${}^{1}$}
\author{Piero \surname{Rettegno}\orcidlink{0000-0001-8088-3517}$^{2}$}
\author{Alessandro \surname{Nagar}${}^{2,3}$}

\affiliation{${}^1$Galileo Galilei Institute for Theoretical Physics, Largo Enrico Fermi, 2, 50125 Firenze, Italy}
\affiliation{${}^2$INFN Sezione di Torino, Via P. Giuria 1, 10125 Torino, Italy} 
\affiliation{${}^3$Institut des Hautes Etudes Scientifiques, 91440 Bures-sur-Yvette, France}
\begin{abstract}
We build upon recent work by Antonelli et al.~[Phys.~Rev.~Lett.~125 (2020) 1, 011103] to obtain, within the effective-one-body (EOB) formalism, and for an arbitrary choice 
of gauge, the third-subleading post-Newtonian (4.5PN) corrections to the spin-orbit conservative dynamics of spin-aligned binaries. 
This is then specialized to: (i) the well-known Damour-Jaranowski-Sch\"afer (DJS) gauge, where the dependence on the angular momentum of the 
gyro-gravitomagnetic functions $(G_S,G_{S_*})$ is removed  and (ii) to an alternative gauge (called anti-DJS gauge, $\overline{\rm DJS}$) that is 
chosen so as to precisely reproduce the Hamiltonian of a spinning test-particle at linear order in the particle spin and keep the full dependence on 
the radial and angular momentum in $(G_S,G_{S_*})$. 
We use these results to extend by one perturbative order, in PN sense, the analytical knowledge of the periastron advance.
After performing a suitable factorization and resummation of $(G_S,G_{S_*})$, the DJS and $\overline{\rm DJS}$ performances  are compared 
via various gauge-invariant quantities at  the EOB last stable circular orbit. 
We eventually find some indications that the $\overline{\rm DJS}$ gauge might be advantageous in the description of the inspiral dynamics of circularized binaries.
\end{abstract}
\date{\today}

\maketitle
\section{Introduction}
\label{sec:intro}

The detection and characterization of gravitational wave (GW) observations from compact binary 
coalescences~\cite{LIGOScientific:2018mvr,LIGOScientific:2020ibl,LIGOScientific:2021djp, Nitz:2021zwj, Olsen:2022pin} 
relies on a precise theoretical prediction of the emitted signal. Highly accurate models of coalescing compact binaries are 
a crucial prerequisite for measuring the properties of their constituent black holes (BHs) and neutron stars, for determining their underlying astrophysical distributions, and for testing General Relativity in the strong-field regime. 

The effective one-body (EOB) formalism~\cite{Buonanno:2000ef,Damour:2000we,Damour:2001tu,Damour:2015isa} 
is currently the only semi-analytical method that allows one to generate accurate waveforms for any type of 
coalescing binary, such as quasi-circular and eccentric BHs~\cite{Akcay:2020qrj,Schmidt:2020yuu,Nagar:2020xsk,Ossokine:2020kjp,Riemenschneider:2021ppj,Gamba:2021ydi,Gamba:2020ljo,Gamba:2021gap,Ramos-Buades:2021adz,Bonino:2022hkj,Placidi:2021rkh,Albanesi:2022xge}, neutron stars~\cite{Lackey:2018zvw,Godzieba:2020bbz,Gamba:2020ljo,Gamba:2020wgg,Breschi:2022ens,Gamba:2022mgx} or mixed binaries~\cite{Matas:2020wab,Gonzalez:2022prs}. 
The EOB method relies on three building blocks: i) a Hamiltonian that describes the conservative part of the dynamics; (ii) the radiation reaction forces, that
describe the back-reaction on the system of the gravitational wave losses, and (iii) a prescription to compute the waveform from the dynamics.

The importance of including spin effects within the EOB Hamiltonian,
\begin{equation} \label{eq:HTEOB}
H_{\rm EOB}=M\sqrt{1+2\nu(\hat{H}_{\rm eff}-1)} \ ,
\end{equation}
where $\hat{H}_{\rm eff}\equiv H_{\rm eff}/\mu$ 
is the (rescaled) EOB Hamiltonian, was pointed out early in Ref.~\cite{Damour:2001tu}.
Here, $M\equiv m_1+m_2$ is the total mass of the binary system,
$\mu\equiv m_1 m_2/M$ is the reduced mass and $\nu\equiv \mu/M$ is the symmetric mass ratio.
The first complete waveform 
model for spinning and precessing coalescing black hole binaries was presented 
in Ref.~\cite{Buonanno:2005xu}. 
Focusing on the conservative dynamics, one separates effects that are
odd in spin (spin-orbit effects) and even in spin. Reference~\cite{Damour:2014sva} (see also~\cite{Balmelli:2015zsa} and~\cite{Khalil:2023kep})
proposed to incorporate even-in-spin effects within a suitable centrifugal radius, assuming that the structure of the
Hamiltonian of a test-particle on a Kerr spacetime is maintained also for comparable mass binaries.

In particular, following Ref.~\cite{Damour:2008qf}, the spin-orbit coupling is encoded into two {\it gyro-gravitomagnetic} functions $(G_S,G_{S_*})$
that, in general, depend on {\it three} variables: the relative separation $R$ between the two objects of masses $(m_1,m_2)$, 
the angular momentum $P_\varphi$ and the relative radial momentum $P_R$. The effective Hamiltonian then reads
\begin{equation}
\label{eq:Heff}
	\hat{H}_{\rm eff} =\hat{H}_{\rm eff}^{\rm orb} + P_\varphi \left(G_S {\bf S} + G_{S_*} {\bf S}_*\right)\,,
\end{equation}
where $\hat{H}_{\rm eff}^{\rm orb}$ is the rescaled orbital effective Hamiltonian, that includes the even-in-spin contributions as mentioned above,
The $({\bf S}, {\bf S}_*)$ are linear combinations of the individual spin vectors $({\bf S}_1,{\bf S}_2)$ and are defined below.

It is always possible to perform a spin-gauge transformation so to suitably 
simplify the analytical expressions of $(G_S,G_{S_*})$. 
In particular, Ref.~\cite{Damour:2008qf} obtained $(G_S,G_{S_*})$ at
next-to-leading-order (NLO, i.e. 2.5PN formal accuracy) imposing a gauge that eliminates the dependence on $P_\varphi$, which is known as Damour-Jaranowski-Schäfer (DJS hereafter) gauge. In Ref.~\cite{Nagar:2011fx} these functions were computed at the 
next-to-next-to-leading-order (N$^2$LO) and given in general, gauge-unfixed form. An analogous calculation was performed in 
Ref.~\cite{Barausse:2011ys}, though in a different gauge, with a result that was shown to be equivalent to the one of~\cite{Nagar:2011fx}.
Recently, Refs.~\cite{Antonelli:2020aeb,Antonelli:2020ybz} extended the $(G_S,G_{S_*})$ knowledge to $\rm N^3LO$ (4.5PN order) 
and Ref.~\cite{Khalil:2021fpm} to $\rm N^4LO$ (5.5PN order). However, both these calculations give explicit expressions in the DJS gauge only.


Here we build upon the procedure outlined in Refs.~\cite{Antonelli:2020aeb,Antonelli:2020ybz} to obtain $(G_S,G_{S_*})$
ab initio without a specific gauge fixing. This tool allows us to explore the performance of other spin gauges at $\rm N^3LO$, 
in particular the one proposed in Ref.~\cite{Rettegno:2019tzh} (see also Refs.~\cite{Barausse:2009xi,Khalil:2023kep} for other possible 
gauge choices), that we shall call anti-DJS ($\overline{\rm DJS}$) hereafter.  Such $\overline{\rm DJS}$ gauge is defined such that the $G_{S_*}$ 
function exactly reduces to the corresponding 
function of a spinning test-body on a Kerr spacetime~\cite{Barausse:2009aa,Barausse:2009xi,Harms:2016ctx}, keeping the 
complete dependence on the momenta. In particular, one of the nice features of this gauge is that the angular momentum always 
appears in the combination $P^2=P_R^2+P_\varphi^2/R^2$.

The paper is organized as follows. In Sec.~\ref{sec:recap} we review the Kerr Hamiltonian and the corresponding EOB one for comparable-mass binaries. 
Sec.~\ref{sec:N3LO_GS_and_GSstar} revolves around the computation in 
full gauge generality of $\big(G_{S}, G_{S_*}\big)$ at $\rm N^3LO$ accuracy in the PN expansion.  The latter is then used in Sec.~\ref{sec:E_and_K} to compute the binding energy and the periastron advance, complete of their spin-orbit component, up to the 4.5PN.
Finally, Sec.~\ref{sec:N3LO_gauge_fixing} is dedicated to explore and compare the gauge fixing choices for $\big(G_{S}, G_{S_*}\big)$.

We use geometric units, $G=c=1$, and dimensionless phase-space variables: the relative separation in the center of mass frame $r\equiv R/M $ 
and the Newtonian potential $u\equiv 1/r$; the orbital phase $\varphi$; the radial momentum $p_r\equiv P_R/\mu$; the orbital 
angular momentum $p_\varphi\equiv P_\varphi / (\mu M)$. We restrict ourselves to the case of spins only along the $z$-direction, $(S_1,S_2)$, 
identified by the orbital angular momentum, and use the following combinations of the individual spins
\begin{align}
	\hat{S}&\equiv \frac{S_1+S_2}{M^2}, \\ 
	\label{eq:Sstar}
	\hat{S}_*&\equiv \frac{1}{M^2}\left(\dfrac{m_2}{m_1}S_1 + \dfrac{m_1}{m_2}S_2\right),
\end{align}
and
\begin{align}
	\tilde{a}_0 \equiv&~ \tilde{a}_1 + \tilde{a}_2 = \hat{S}+\hat{S}_*, \\
	\tilde{a}_{12} \equiv&~ \tilde{a}_1 - \tilde{a}_2 = \frac{\hat{S}-\hat{S}_*}{X_{12}},
\end{align}
where $\tilde{a}_i \equiv S_i/(m_i M)$ are the rescaled spin magnitudes of the two bodies and $X_{12} \equiv (m_1-m_2)/M=\sqrt{1-4\nu}$.


\section{The EOB Hamiltonian for spin-aligned binaries}
\label{sec:recap}
In this section, we recall the structure of the EOB Hamiltonian for spin-aligned binaries, as it was first defined in Ref.~\cite{Damour:2014sva}. 
This is a $\nu$-dependent deformation of the Hamiltonian 
of a spinning particle moving in a Kerr background~\cite{Damour:2007nc,Barausse:2009aa,Barausse:2009xi,Barausse:2011ys,Bini:2015xua}, which we will remind below.

\subsection{Hamiltonian of a spinning particle in a Kerr background}
\label{subsec:spinningparticle}

We start by recalling the Hamiltonian of a spinning particle
orbiting around a Kerr BH~\cite{Damour:2007nc,Barausse:2009aa,Barausse:2009xi,Barausse:2011ys,Bini:2015xua}.
In this scenario, when $m_1 \gg m_2$ ($\nu \ll 1$), we can use $M$ as the central BH mass and $\mu$ as the mass of the test particle.
The spin variables entering the spin-orbit interaction in the comparable mass case [see Eq.~\eqref{eq:Heff}] reduce to the individual dimensionless spins as $\hat{S} \rightarrow \hat{a}_1$ and 
$\hat{S}_* \rightarrow \hat{a}_2$, where $\hat{a}_i \equiv \tilde{a}_i/M$.

In the case of spin vector aligned with the orbital angular momentum, the motion of a spinning particle around a Kerr BH can be described by the compact Hamiltonian~\cite{Damour:2014sva,Bini:2015xua}
\begin{align}
	\label{eq:kerrspH}
	\hat{H}^{K} &\equiv \dfrac{H^K}{\mu} = \sqrt{ A^{K} \left[ 1+p_\varphi^2 \big(u^K_c\big)^2 + \dfrac{A^{K}}{D^{K}} p^2_r \right]}  \cr &+   \left( G_S^{K} \hat{a}_{1} + G_{S_*}^{K} \hat{a}_{2} \right) p_\varphi\,,
\end{align}
where we introduced $u^K_c\equiv 1/r^K_c $, 
the inverse of the centrifugal radius $r^K_c$, defined as 
\begin{equation}
	\label{eq:kerr_rc}
	\big(r^{K}_c\big)^2 
	\equiv r^2+\tilde{a}_1^2\left(1+\dfrac{2}{r}\right).
\end{equation}
The $A^{K}$ and $D^{K}$ terms are the Kerr metric potentials and read
\begin{align}
	\label{eq:kerr_A}
	A^{K} &=  \left( 1 - 2 u^K_c \right) \dfrac{1 + 2 u^K_c}{1+2u},
	\\
	\label{eq:kerr_B}
	D^{K} & = \dfrac{(u_c^{K})^2}{u^2}.
\end{align}
The $G_S^{K}$ and $G_{S_*}^{K}$ factors parametrize the spin-orbit interaction of the two-body system \cite{Bini:2015xua}. They are called gyro-gravitomagnetic functions and read 
\begin{align}
	\label{eq:kerr_GS}
	G_S^{K} & = 2\, u\, (u_c^{K})^2,\\
	\label{eq:kerr_GSstar}
	G_{S_*}^{K}&=(u_c^{K})^2  \left\{ \sqrt{\frac{A^K}{W^K}}\left[ 1 - \frac{u^2\big(u_c^K\big)' }{\big(u_c^K\big)^2} \sqrt{\frac{A^K}{D^K \vphantom{W^K}}} \right] \right. \cr&- \frac{u^2 \, \left(A^K\right)'}{2 u_c^K\left(1+\sqrt{W^K}\right) \sqrt{D^K}} \left. \vphantom{\sqrt{\frac{A^K}{W^K}}} \right\},
\end{align}
where the primes denote partial derivatives w.r.t. $u$ and $W^K$ is the factor entering the square root of Eq.~\eqref{eq:kerrspH}, namely
\begin{align}
	W^K \equiv 1 + p_\varphi^2(u^{K}_c)^2 + \dfrac{A^{K}}{D^{K}}p^2_r.
\end{align}

\subsection{Hamiltonian for comparable-mass spin-aligned binaries}
\label{subsec:TEOB_Hamiltonian}

Following Ref.~\cite{Damour:2014sva}, we write an effective Hamiltonian for binary systems with arbitrary mass ratios [see Eqs.~\eqref{eq:HTEOB} and \eqref{eq:Heff}] by mimicking the structure of the Kerr Hamiltonian 
and allowing for $\nu$-dependent deformations in each of its building blocks.
The rescaled effective Hamiltonian then reads
\begin{align}
	\label{eq:HeffTEOB}
	\hat{H}_{\rm eff} =&~ \sqrt{A \left[ 1+p_\varphi^2 u_c^2 + 
	Q	\right] + p_{r_*}^2}\cr&+ \left( G_S \hat{S} + G_{S_*} \hat{S}_* \right) p_\varphi,
\end{align}
where we introduced the tortoise-coordinate radial momentum $p_{r_*} = A/\sqrt{D} \, p_r $.
The first term of Eq.~\eqref{eq:HeffTEOB} takes into account even-in-spin interactions (including the spin-independent ones), 
while the second one determines the spin-orbit interaction.

Even-in-spin effects are fully encoded in $u_c\equiv 1/r_c$, where $r_c$ is the EOB centrifugal radius~\cite{Damour:2014sva}. This is
defined according to the structure of $r_c^K$, Eq.~\eqref{eq:kerr_rc}, and reads
\begin{equation}
	\label{rcTEOB}
	r_c^2 \equiv r^2 + \tilde{a}_0^2\left(1+\dfrac{2}{r}\right) + \dfrac{\delta a^2}{r},
\end{equation}
with the NLO spin-spin term given by
\begin{equation}
	\delta a^2 \equiv -\dfrac{1}{8} \Bigg\{9\, \tilde{a}_0^2 
	+ \left(1 + 4\nu\right) \tilde{a}_{12}^2 - 10 X_{12}\, \tilde{a}_0\, \tilde{a}_{12}	\Bigg\}.
\end{equation}

The potentials $A$ and $D$ in the Hamiltonian are a $\nu$-deformed version of $A^K$ and $D^K$ and read 
\begin{align}
	\label{eq:Aeob}
	A &= A_{\rm orb} \left(u_c\right) \dfrac{1 + 2 u_c}{1+2 u}, 
	\\
		\label{eq:Deob}
	D & = D_{\rm orb} (u_c) \dfrac{u_c^2}{u^2} ,
\end{align}
where $A_{\rm orb}$ and $D_{\rm orb}$ are the nonspinning EOB metricc potentials (evaluated as functions of $u_c$). 
The $Q$ potential instead collects all the extra-geodesic (\textit{i.e.}~more than quadratic in the momenta) corrections entering the effective Hamilton-Jacobi equation from 3PN onward.
These are not present in the Kerr Hamiltonian, Eq.~\eqref{eq:kerrspH}, and are accordingly all proportional to $\nu$ (see below).

In particular, at the 4PN accuracy we will need in the following,  we have \cite{Damour:2015isa}
\begin{subequations} \label{eq:4PN_potentials}
	\begin{align}
			\label{eq:A4PN}
			&A^{\rm 4PN}_{\rm orb}(u_c) = 1 - 2 u_c + 2 \nu u_c^3 + \left( \dfrac{94}{3} - \dfrac{41 \pi^2}{32} \right) \nu u_c^4 \notag \\ &\hspace{1cm} + \left( a_5^c + \dfrac{64}{5} \nu \log{u_c} \right) u_c^5\,, \\
			& D^{\rm 4PN}_{\rm orb}(u_c) = 1 - 6 \nu u_c^2 - 2\left(26-3 \nu \right)\nu u_c^3 \notag \\ &\hspace{1cm}+ \bigg( d_4^c + \dfrac{592}{15} \nu \log{u_c} \bigg) u_c^4\,,\\
			&Q^{\rm 4PN}(p_r,u_c) =  \big[2 \nu (4-3\nu)u_c^2 +q_{43} u_c^3 \big]p_{r}^4 \notag \\ &\hspace{1cm}+ q_{62}u_c^2 p_r^6\,,
	\end{align}
\end{subequations}
where 
\begin{subequations}
	\begin{align}
		&a_5^c = \bigg( \dfrac{2275 \pi^2}{512} - \dfrac{4237}{60} + \dfrac{128}{5} \gamma_{\rm E} + \dfrac{256}{5} \log{2} \bigg) \nu \notag \\
		&\hspace{1cm}+ \bigg( \dfrac{41 \pi^2}{32} - \dfrac{221}{6} \bigg) \nu^2, \\
		&d_4^c = \bigg(-\frac{533}{45}+\frac{1184 \gamma_{\rm E} }{15}-\frac{23761 \pi ^2}{1536}-\frac{6496 \log
			2}{15} \notag \\
		&\hspace{1cm}+\frac{2916 \log 3}{5}\bigg) \nu +\bigg(-260+\frac{123 \pi ^2}{16}\bigg)
		\nu ^2,\\
		&q_{43} = \bigg(-\frac{5308}{15}+\frac{496256 \log 2}{45}-\frac{33048 \log 3}{5}\bigg) \nu \notag \\
		&\hspace{1cm}-83
		\nu ^2+10 \nu ^3, \\
		&q_{62} =  \bigg(-\frac{827}{3}-\frac{2358912 \log 2}{25}+\frac{1399437 \log
			3}{50}\notag \\
		&\hspace{1cm}+\frac{390625 \log 5}{18}\bigg)\nu -\frac{27 \nu ^2}{5}+6 \nu ^3,
	\end{align}
\end{subequations}
with $\gamma_{\rm E}$ being Euler's constant.
Note that here $Q$ is expressed in a gauge where its dependence on $p_\varphi$ is removed \cite{Damour:1999cr}.

The second term of Eq.~\eqref{eq:HeffTEOB} is written in terms of the two effective gyro-gravitomagnetic functions $\big(G_S,G_{S_*}\big)$, which generalize to general mass ratios the test-mass expressions $\big(G^K_S,G^K_{S_*}\big)$ and determine the strength of the spin-orbit coupling during the binary evolution. 
These will be the main focus of this work.

\section{Spin-orbit part of the EOB Hamiltonian up to $\rm \bold{N^3LO}$ in full generality}
\label{sec:N3LO_GS_and_GSstar}
In this section we derive for the first time the gauge-unfixed expressions for the functions $\big(G_S,G_{S_{*}}\big)$ up to $\rm N^3LO$. 
We closely follow the procedure of Refs.~\cite{Antonelli:2020ybz,Antonelli:2020aeb}, although avoiding to specify a chosen gauge from the start.

\subsection{Computation setup}
\label{subsec:computation_setup}

The starting point of our computation is the most general PN ansatz that is dimensionally allowed for $G_{S}$ and $G_{S_*}$. 
Explicitly, considering a set of dimensionless $\nu$-dependent coefficients $g_{n}^{\text{N}^m\text{LO}}$, the 
generic ansatz for $G_{S}$ reads\footnote{For simplicity, we fix from the start the leading order to its known value.}
\begin{widetext}
\begin{align}
	\label{eq:generalGS}
	&G^{\rm gen}_{S} = u^3\left[ 2  +\dfrac{1}{c^2}\left( 
	 g_{1}^{\rm NLO} p^2  +  g_{2}^{\rm NLO} p_r^2+  g_{3}^{\rm NLO} u 
	\right)+ \dfrac{1}{c^4}\left(
	 g_{1}^{\rm N^2LO} p^4+  g_{2}^{\rm N^2LO}p^2 p_r^2+  g_{3}^{\rm N^2LO} p^2 u + g_{4}^{\rm N^2LO} p_r^4 \right. \right. \cr & \left. \left. \hspace{0.5cm} +  g_{5}^{\rm N^2LO} p_r^2 u+  g_{6}^{\rm N^2LO}u^2
	 \right)+ \dfrac{1}{c^6} \left(
	   g_{1}^{\rm N^3LO} p^6 +  g_{2}^{\rm N^3LO}p^4 p_r^2 +  g_{3}^{\rm N^3LO} p^4 u+  g_{4}^{\rm N^3LO} p^2 p_r^4+  g_{5}^{\rm N^3LO} p^2 p_r^2 u \right. \right. \cr & \left. \left. \hspace{0.5cm}+  g_{6}^{\rm N^3LO} p^2 u^2 +  g_{7}^{\rm N^3LO} p_r^6  + g_{8}^{\rm N^3LO} p_r^4 u +  g_{9}^{\rm N^3LO} p^2 u^2+  g_{10}^{\rm N^3LO} u^3
	 \right)\right],
\end{align}
\end{widetext} 
while $G^{\rm gen}_{S_*}$, the corresponding ansatz for $G_{S_*}$, has the same structure of $G^{\rm gen}_{S}$ with a 3/2 replacing the 2 at leading order and a different, independent set of coefficients $g_{*n}^{\text{N}^m\text{LO}}$.
In writing Eq.~\eqref{eq:generalGS} we introduced the dimensionless total momentum $p^2=p_r^2+p_\varphi^2 u^2$ and singled out each PN order beyond the leading one by restoring powers of $1/c$. We also specify that, as we are here interested in contributions linear in the spin, the difference between $u_c$ and $u$ is not relevant, therefore each quantity is written in terms of the latter.

It is important to notice that, at this stage, $G^{\rm gen}_{S}$ and  $G^{\rm gen}_{S_*}$ are still devoid of any physical meaning. The actual $\big(G_S,G_{S_*}\big)$ we are looking for must reproduce the spin-orbit part of the dynamics of a spinning binary, a condition that is satisfied only if certain relations hold between the coefficients $g_{n}^{\text{N}^m\text{LO}}$ and, separately, $g_{*n}^{\text{N}^m\text{LO}}$. 

Our source of dynamical information is the scattering angle of a pair of gravitationally-interacting spinning bodies, which is known to encode in gauge-invariant form the entire local-in-time conservative dynamics of an aligned-spin binary \cite{Damour:2016gwp,Damour:2017zjx}. We refer in particular to the contributions proportional to $\hat{S}$ and $\hat{S}_*$ in Eq.~(4.38) of Ref.~\cite{Antonelli:2020ybz}\footnote{Ref.~\cite{Antonelli:2020ybz} denotes our $\hat{S}$ and $\hat{S}_*$ respectively as $a_b$ and $a_t$.}, which gives the scattering angle of two spinning bodies at the third subleading PN order, and thus we dub hereafter as $\chi_{\rm 3PN}$.

The basic idea is to compute the scattering angle from an effective Hamiltonian of the type \eqref{eq:HeffTEOB} whose spin-orbit part is written in terms of $\big(G^{\rm gen}_S,G^{\rm gen}_{S_*}\big)$, and then match the result to $\chi_{\rm 3PN}$. 

\subsection{The scattering angle from the effective Hamiltonian}
	\label{subsec:effective_scattering_angle}
	
The scattering angle associated to the effective Hamiltonian $\hat{H}_{\rm eff}$ is given by the following integral
\begin{equation}
	\label{eq:chi_eff_integral}
	\chi (\hat{E}_{\rm eff}, p_\varphi) \equiv -\pi-2\int^{u_{\rm max}}_{0}  \dfrac{du}{u^2} \, \dfrac{\partial }{\partial p_\varphi}p_r(\hat{E}_{\rm eff},p_\varphi,u),
\end{equation}
where $p_r(\hat{E}_{\rm eff},p_\varphi,u)$ is the radial momentum obtained by inverting perturbatively, in PN sense, the energy conservation 
equation $\hat{H}_{\rm eff} = \hat{E}_{\rm eff}$ at fixed $\hat{E}_{\rm eff}$~\footnote{The (rescaled) effective energy $\hat{E}_{\rm eff}$ is equal to the relativistic Lorentz factor of the system $\gamma \equiv - u_1 \cdot u_2$, where $u_i$ 
is the 4-velocity of the body $i$ in the binary.}, and $u_{\rm max}$ is the largest real root of $p_r(\hat{E}_{\rm eff},p_\varphi,u)=0$.

By naively expanding the integrand in Eq.~\eqref{eq:chi_eff_integral} one would generate a series of formally divergent integrals, with the additional 
degree of complexity that the upper bound $u_{\rm max}$ is itself given by a PN expansion, $u_{\rm max}= p_\infty/p_\varphi + \mathcal{O}(1/c^2)$, where $p_\infty \equiv \sqrt{\hat{E}_{\rm eff}^2-1}$ . Nevertheless, this integral is easily computed following the procedure of Ref.~\cite{Damour:1988mr}, that we outline here for completeness.
\begin{itemize}
	\item[(i)] We introduce the integration variable $z=p_\varphi u/p_\infty$, such that $z_{\rm max}= 1+ \mathcal{O}(1/c^2)$.
	\item[(ii)] We expand the integrand in $1/p_\varphi$ up to $1/p_\varphi^4$ and in $1/c$ up to $1/c^8$.\footnote{At leading order, the PN expansion of the scattering angle is proportional to $1/c^2$, thus  $1/c^8$ corresponds to the third subleading PN order.}
	\item[(iii)] We ignore any PN correction in $z_{\rm max}$ and regularize the divergent integrals by taking the \textit{Hadamard partie finie} ($\rm Pf$), \text{\textit{i.e.}}
	\begin{equation}
		\int_0^{1+\mathcal{O}(G/c^2)} dz \,\rightarrow \,{\rm Pf} \int_0^{1} dz.
	\end{equation}
\item[(iv)] After the previous steps, each integral in the expansion has the structure
\begin{equation} \label{eq:integral_structure}
	{\rm Pf} \int_0^{1} dz \, (1-z^2)^{-1/2-n} \,z^m,
\end{equation}
which is actually the ${\rm Pf}$ of an Euler Beta function, as can be seen explicitly by changing the integration variable to $t=z^2$. Such a finite part can be simply evaluated via analytical continuation and accordingly Eq.~\eqref{eq:integral_structure} is equivalent to
\begin{equation}
	\lim_{\eta \to 0} \int_0^1 dz \, (1-z^2)^{-1/2+\eta-n} \,z^m.
\end{equation}
\end{itemize}

During this procedure we need also to be careful about the difference between the canonical angular momentum $p_\varphi$ appearing in the effective Hamiltonian and the ``covariant" one that appears in $\chi_{\rm 3PN}$ in the form $p^{\rm cov}_\varphi = b\, p_\infty/(G M \hat{E})$.
Here, $b$ is the covariant impact parameter and $\hat{E} \equiv E/M$ is the rescaled total energy of the system, related to the effective one by the usual EOB energy map [Eq.~\eqref{eq:HTEOB}], namely
\begin{equation}
	\hat{E} \equiv \sqrt{1+2 \nu \left( \hat{E}_{\rm eff}- 1 \right)}\,.
\end{equation}
The two angular momenta are related by \cite{Vines:2017hyw}
\begin{equation}
	p_\varphi = p_\varphi^{\rm cov}+\frac{\hat{E} -1}{ 2 \nu \, c^2 }
	\left(\hat{S}+\hat{S}_*-\frac{\hat{S}-\hat{S}_*}{\hat{E}}\right).
\end{equation}

Starting from an effective Hamiltonian with a 4PN orbital part [that is with the 4PN EOB potentials of Eq.~\eqref{eq:4PN_potentials}] 
and with the general gyro-gravitomagnetic functions $\big(G^{\rm gen}_S,G^{\rm gen}_{S_*}\big)$ in the spin-orbit term, 
the result we get for the scattering angle has the following three-component structure 
\begin{align}
	&\chi 
	= \chi^{\rm orb} (\hat{E}_{\rm eff}, p_\varphi) + \hat{S}\, \chi^{\rm SO}_S (\hat{E}_{\rm eff}, p_\varphi,g_{n}^{{\rm N}^m{\rm LO}})
	\cr&\hspace{0.5cm}+ \hat{S_*}\, \chi^{\rm SO}_{S_*} (\hat{E}_{\rm eff}, p_\varphi ,g_{ *n}^{{\rm N}^m{\rm LO}} ).
\end{align}
Here the spin-free part  $ \chi^{\rm orb}$ coincides with that of $\chi_{\rm 3PN}$ (see the first three lines in Eq.~(4.38) of Ref.~\cite{Antonelli:2020ybz}) while the spin-orbit parts $(\chi^{\rm SO}_S,\chi^{\rm SO}_{S_*})$ can be matched to the corresponding ones in $\chi_{\rm 3PN}$ by constraining accordingly the coefficients of our starting ansatz, on which they both depend. 

\subsection{ $\rm N^3LO$-accurate $G_{S}$ and $G_{S_*}$ in a generic spin-gauge}
\label{subsec:gauge_unfixed_GS_and_GSstar}
Each of the two spin-orbit component of $\chi$, once matched to their counterparts in $\chi_{\rm 3PN}$, give rise to 9 $\nu$-dependent relations among the coefficients of $G^{\rm gen}_{S}$ and $G^{\rm gen}_{S_*}$, respectively; all of them are explicitly given in Appendix \ref{app:coeff_rel_matching}. The resulting gauge-unfixed expressions for $G_{S}$ and $G_{S_*}$ have a total of 10 residual gauge coefficients each and, introducing the notation $g_{X}\equiv G_{X}/u^3$ for $X=S,S_{*}$, they read
\begin{widetext}
\begin{subequations}
\begin{align} \label{eq:gauge_unfixed_GS}
	&g_S=
	2 + \frac{1}{c^2} \bigg[
	g_2^{\rm NLO} p_r^2+ \bigg(-\frac{g_2^{\rm NLO}}{3}-\frac{9 \nu }{8}\bigg)p^2+
	\bigg(\frac{g_2^{\rm NLO}}{3}+\frac{\nu }{2}\bigg)u
	\bigg]+ \frac{1}{c^4} \bigg[
	g_2^{\rm N^2LO} p^2 p_r^2+g_4^{\rm N^2LO} p_r^4+g_5^{\rm N^2LO} p_r^2 u\cr &\hspace{0.5cm}+ \bigg(\frac{g_2^{\rm N^2LO}}{2}+\frac{g_2^{\rm NLO}}{4}+\frac{9
		g_4^{\rm N^2LO}}{20}-\frac{g_5^{\rm N^2LO}}{4}-\frac{33 \nu }{16}-\frac{5 \nu
		^2}{4}\bigg)p^2
	u+ \bigg(-\frac{g_2^{\rm N^2LO}}{6}+\frac{3
		g_2^{\rm NLO}}{4}-\frac{g_4^{\rm N^2LO}}{4}+\frac{g_5^{\rm N^2LO}}{4}\cr &\hspace{0.5cm}-\frac{119 \nu
	}{16}+\frac{\nu ^2}{4}\bigg)u^2+
	\bigg(-\frac{g_2^{\rm N^2LO}}{3}-\frac{g_4^{\rm N^2LO}}{5}+\frac{\nu }{8}+\frac{7 \nu
		^2}{8}\bigg)p^4
	\bigg]+ \frac{1}{c^6} \bigg[
	g_2^{\rm N^3LO} p^4 p_r^2+g_4^{\rm N^3LO} p^2 p_r^4+g_7^{\rm N^3LO}
	p_r^6\cr &\hspace{0.5cm}+g_5^{\rm N^3LO} p^2 p_r^2 u+g_8^{\rm N^3LO} p_r^4 u+g_9^{\rm N^3LO}
	p_r^2 u^2+\bigg(\frac{2 g_2^{\rm N^2LO}}{3}-\frac{7
		g_2^{\rm N^3LO}}{15}+\frac{17 g_2^{\rm NLO}}{30}+\frac{63 g_4^{\rm N^2LO}}{200}-\frac{3
		g_4^{\rm N^3LO}}{5}\cr &\hspace{0.5cm}+\frac{g_5^{\rm N^2LO}}{10}+\frac{7 g_5^{\rm N^3LO}}{20}-\frac{5
		g_7^{\rm N^3LO}}{8}+\frac{11 g_8^{\rm N^3LO}}{40}-\frac{g_9^{\rm N^3LO}}{5}-\frac{1231 \nu
	}{80}+\frac{2 g_2^{\rm NLO} \nu }{5}+\frac{431 \nu ^2}{40}-\frac{11 \nu
		^3}{8}\bigg)p^2 u^2 \cr &\hspace{0.5cm}+
	\bigg(-\frac{g_2^{\rm N^3LO}}{3}-\frac{g_4^{\rm N^3LO}}{5}-\frac{g_7^{\rm N^3LO}}{7}+\frac{
		\nu }{128}-\frac{9 \nu ^2}{32}-\frac{95 \nu ^3}{128}\bigg)p^6+
	\bigg(-\frac{g_2^{\rm N^2LO}}{2}+\frac{2 g_2^{\rm N^3LO}}{15}+\frac{101
		g_2^{\rm NLO}}{60}-\frac{21 g_4^{\rm N^2LO}}{25}\cr &\hspace{0.5cm}+\frac{g_4^{\rm N^3LO}}{5}+\frac{13
		g_5^{\rm N^2LO}}{20}-\frac{g_5^{\rm N^3LO}}{10}+\frac{g_7^{\rm N^3LO}}{4}-\frac{3
		g_8^{\rm N^3LO}}{20}+\frac{g_9^{\rm N^3LO}}{5}-\frac{28331 \nu }{720}-\frac{7
		g_2^{\rm NLO} \nu }{5}+\frac{241 \pi ^2 \nu }{192}-\frac{123 \nu ^2}{20}+\frac{\nu
		^3}{8}\bigg)u^3 \cr &\hspace{0.5cm}+ \bigg(\frac{g_2^{\rm N^2LO}}{3}+\frac{2 g_2^{\rm N^3LO}}{3}+\frac{3
		g_4^{\rm N^2LO}}{8}+\frac{3 g_4^{\rm N^3LO}}{5}-\frac{g_5^{\rm N^3LO}}{4}+\frac{29
		g_7^{\rm N^3LO}}{56}-\frac{g_8^{\rm N^3LO}}{8}+\frac{\nu }{2}+\frac{21 \nu
		^2}{8}+\frac{31 \nu ^3}{16}\bigg)p^4 u
	\bigg],  
	\\
	\cr
	 \label{eq:gauge_unfixed_GSstar}
	&g_{S_*}=  \frac{3}{2} + \frac{1}{c^2}\bigg[
	\bigg(-\frac{5}{8}-\frac{g_{*2}^{\rm NLO}}{3}\bigg) p^2+g_{*2}^{\rm NLO}
	p_r^2+\bigg(-\frac{1}{2}+\frac{g_{*2}^{\rm NLO}}{3}\bigg) u
	\bigg] 
	+ \frac{1}{c^4}\bigg[
	\bigg(\frac{7}{16}-\frac{g_{*2}^{\rm N^2LO}}{3}-\frac{g_{*4}^{\rm N^2LO}}{5}\bigg)
	p^4\cr &\hspace{0.5cm}+g_{*2}^{\rm N^2LO} p^2 p_r^2+g_{*4}^{\rm N^2LO}
	p_r^4+\bigg(\frac{9}{16}+\frac{g_{*2}^{\rm N^2LO}}{2}+\frac{g_{*2}^{\rm NLO}}{4}
	+\frac{9 g_{*4}^{\rm N^2LO}}{20}-\frac{g_{*5}^{\rm N^2LO}}{4}\bigg) p^2
	u+g_{*5}^{\rm N^2LO} p_r^2
	u+\bigg(-\frac{13}{16}-\frac{g_{*2}^{\rm N^2LO}}{6}\cr &\hspace{0.5cm}+\frac{3
		g_{*2}^{\rm NLO}}{4}-\frac{g_{*4}^{\rm N^2LO}}{4}+\frac{g_{*5}^{\rm N^2LO}}{4}\bigg)
	u^2
	\bigg] +  \frac{1}{c^6}\bigg[
	\bigg(-\frac{45}{128}-\frac{g_{*2}^{\rm N^3LO}}{3}-\frac{g_{*4}^{\rm N^3LO}}{5}-\frac{g_{*7}^{\rm N^3LO}}{7}\bigg) p^6+g_{*2}^{\rm N^3LO} p^4 p_r^2+g_{*4}^{\rm N^3LO} p^2
	p_r^4\cr &\hspace{0.5cm}+g_{*7}^{\rm N^3LO}
	p_r^6+\bigg(-\frac{5}{8}+\frac{g_{*2}^{\rm N^2LO}}{3}+\frac{2
		g_{*2}^{\rm N^3LO}}{3}+\frac{3 g_{*4}^{\rm N^2LO}}{8}+\frac{3
		g_{*4}^{\rm N^3LO}}{5}-\frac{g_{*5}^{\rm N^3LO}}{4}+\frac{29
		g_{*7}^{\rm N^3LO}}{56}-\frac{g_{*8}^{\rm N^3LO}}{8}\bigg) p^4 u\cr &\hspace{0.5cm}+g_{*5}^{\rm N^3LO} p^2
	p_r^2 u+g_{*8}^{\rm N^3LO} p_r^4 u+\bigg(\frac{51}{80}+\frac{2
		g_{*2}^{\rm N^2LO}}{3}-\frac{7 g_{*2}^{\rm N^3LO}}{15}+\frac{17
		g_{*2}^{\rm NLO}}{30}+\frac{63 g_{*4}^{\rm N^2LO}}{200}-\frac{3
		g_{*4}^{\rm N^3LO}}{5}+\frac{g_{*5}^{\rm N^2LO}}{10}+\frac{7
		g_{*5}^{\rm N^3LO}}{20}\cr &\hspace{0.5cm}-\frac{5 g_{*7}^{\rm N^3LO}}{8}+\frac{11
		g_{*8}^{\rm N^3LO}}{40}-\frac{g_{*9}^{\rm N^3LO}}{5}\bigg) p^2 u^2+g_{*9}^{\rm N^3LO}
	p_r^2 u^2+\bigg(-\frac{121}{80}-\frac{g_{*2}^{\rm N^2LO}}{2}+\frac{2
		g_{*2}^{\rm N^3LO}}{15}+\frac{101 g_{*2}^{\rm NLO}}{60}-\frac{21
		g_{*4}^{\rm N^2LO}}{25}\cr &\hspace{0.5cm}+\frac{g_{*4}^{\rm N^3LO}}{5}+\frac{13
		g_{*5}^{\rm N^2LO}}{20}-\frac{g_{*5}^{\rm N^3LO}}{10}+\frac{g_{*7}^{\rm N^3LO}}{4}-\frac
	{3 g_{*8}^{\rm N^3LO}}{20}+\frac{g_{*9}^{\rm N^3LO}}{5}\bigg) u^3
	\bigg].
\end{align}
\end{subequations}
\end{widetext}
These expressions represent the main result of the paper and extend up to $\rm N^3LO$ the $\rm N^2LO$-accurate gauge-unfixed gyro-gravitomagnetic functions first obtained in Ref.~\cite{Nagar:2011fx} and provided in Eq.~(29) therein. To explicitly see the correspondence at $\rm NLO$ and $\rm N^2LO$ between the results of Ref.~\cite{Nagar:2011fx} and those given above one has however to make the following shifts on the gauge coefficients:
\begin{subequations}
	\begin{align}
		&g_2^{\rm NLO}\to -3 a-\frac{9 \nu }{2}, \\ &g_2^{\rm N^2LO}\to \frac{3 a}{2}+3 \beta -3 \gamma +\frac{9 \nu }{4}-\frac{39 \nu ^2}{16}, \\
		&g_4^{\rm N^2LO}\to -5 \beta +\frac{135 \nu ^2}{16},\\
		&g_5^{\rm N^2LO}\to 6 a-4 \alpha -3 \beta -2 \gamma +\frac{35 \nu }{4}-\frac{3 \nu \
			^2}{16},
		\end{align}
\end{subequations}
and
\begin{subequations}
	\begin{align}
		&g_{*2}^{\rm NLO}\to -3 b-\frac{15 \nu }{4}, \\ 
		&g_{*2}^{\rm N^2LO}\to \frac{3 b}{2}+3 \zeta -3 \eta +\frac{57 \nu }{16}-\frac{21 \nu ^2}{8}, \\
		&g_{*4}^{\rm N^2LO}\to6 b-4 \delta -3 \zeta -2 \eta +\frac{5}{4}+\frac{109 \nu }{8}+\frac{3 \nu ^2}{4}, \\
		&g_{*5}^{\rm N^2LO}\to -5 \zeta +\frac{15 \nu ^2}{2}.
	\end{align}
\end{subequations}

\section{Gauge invariant quantities}
\label{sec:E_and_K}
In order to check our general $\rm N^3LO$  results for $\big(G_S,G_{S_{*}}\big)$, we compute here two gauge invariant quantities, complete of their spin orbit part: the effective binding energy and the fractional advance of the periastron per radial period, both obtained in the \emph{adiabatic approximation}, that is by approximating the dynamics as a sequence of circular orbits, with 4.5PN accuracy. 

The (rescaled) binding energy is defined as
\begin{equation}
	\label{eq:Eb_def}
		E_b \equiv \hat{H}_{\rm EOB}^{\rm circ}(x) - \frac{1}{\nu},
\end{equation}
where $x\equiv \Omega^{2/3}$ and $\hat{H}_{\rm EOB}^{\rm circ}(x)$ is the $\mu$-rescaled version of the EOB Hamiltonian \eqref{eq:HTEOB} in the limit $p_{r_*} \to 0$, with $p_\varphi$ and $u$ replaced by their 4.5PN circular expansions in terms of $x$. These can be determined from Eq.~\eqref{eq:HTEOB}, first by obtaining the circular expansion of $p_\varphi$ in powers of $u$, as it can be done by solving perturbatively the equation
\begin{equation}
	0=\dot{p}_{r_*} = \frac{\partial \hat{H}_{\rm EOB}(p_{r_*},p_\varphi,u)}{\partial r}\Big|_{p_{r_*}=0}\,,
\end{equation}
for $p_\varphi(u)$, and then by computing the circular expansion of $u$ in powers of $x$ via the perturbative inversion of the equation
\begin{equation}
	x \equiv \Omega_\varphi^{2/3}= \bigg[\frac{\partial \hat{H}_{\rm EOB}(p_{r_*},p_\varphi,u)}{\partial p_\varphi}\Big|_{p_{r_*}=0}\bigg]^{2/3}.
\end{equation}
Using our $\big(G_S,G_{S_{*}}\big)$ in the EOB Hamiltonian of Eqs.~\eqref{eq:HTEOB}-\eqref{eq:HeffTEOB}, the resulting circular expansions are
\begin{widetext}
	\begin{align}
		\label{eq:pphi}
		&p_\varphi= \frac{1}{\sqrt{x}}+\sqrt{x} \bigg(\frac{3}{2}+\frac{\nu }{6}\bigg)+x^{3/2}
		\bigg(\frac{27}{8}-\frac{19 \nu }{8}+\frac{\nu ^2}{24}\bigg)+x^{5/2}
		\bigg[\frac{135}{16}+\bigg(-\frac{6889}{144}+\frac{41 \pi ^2}{24}\bigg) \nu
		+\frac{31 \nu ^2}{24}+\frac{7 \nu ^3}{1296}\bigg]\cr&\hspace{0.5cm}+x^{7/2}
		\bigg[\frac{2835}{128}+\bigg(\frac{98869}{5760}-\frac{128 \gamma_{\rm E} }{3}-\frac{6455 \pi
			^2}{1536}-\frac{256 \log 2}{3}-\frac{64 \log x}{3}\bigg) \nu
		+\bigg(\frac{356035}{3456}-\frac{2255 \pi ^2}{576}\bigg) \nu ^2-\frac{215 \nu
			^3}{1728}-\frac{55 \nu ^4}{31104}\bigg]\cr&\hspace{0.5cm}-x \bigg(\frac{10}{3} \hat{S}+\frac{5
			\hat{S}_*}{2}\bigg)+x^2 \bigg[\hat{S}
		\bigg(-7+\frac{217 \nu }{72}\bigg)+\hat{S}_* \bigg(-\frac{21}{8}+\frac{35 \nu }{12}\bigg)\bigg]+x^3 \bigg[\hat{S}
		\bigg(-\frac{81}{4}+\frac{633 \nu }{16}-\frac{7 \nu ^2}{8}\bigg)\cr&\hspace{0.5cm}+\hat{S}_*
		\bigg(-\frac{81}{16}+\frac{117 \nu }{4}-\frac{15 \nu ^2}{16}\bigg)\bigg]+x^4
		\bigg\{\hat{S}
		\bigg[-\frac{495}{8}+\bigg(\frac{216469}{1152}+\frac{319 \pi ^2}{192}\bigg) \nu
		-\frac{21769 \nu ^2}{288}-\frac{2915 \nu ^3}{31104}\bigg]\cr&\hspace{0.5cm}+\hat{S}_* \bigg(-\frac{1485}{128}+\frac{6215 \nu }{64}-\frac{12199 \nu
			^2}{192}-\frac{275 \nu ^3}{2592}\bigg)\bigg\},
		\\ \cr
		\label{eq:u}
		&u = x-\frac{ \nu }{3}x^2+\frac{5  \nu }{4}x^3+x^4 \bigg[\bigg(\frac{1585}{72}-\frac{41 \pi
			^2}{48}\bigg) \nu -\frac{7 \nu ^2}{4}+\frac{\nu ^3}{81}\bigg]+x^5
		\bigg[\bigg(-\frac{153211}{2880}+\frac{64 \gamma_{\rm E} }{3}+\frac{11375 \pi
			^2}{3072}+\frac{128 \log 2}{3}\cr&\hspace{0.5cm}+\frac{32 \log x}{3}\bigg) \nu
		+\bigg(-\frac{31777}{432}+\frac{1681 \pi ^2}{576}\bigg) \nu ^2+\frac{3 \nu
			^3}{4}+\frac{\nu ^4}{243}\bigg]+x^{5/2} \bigg(\frac{2
			\hat{S}}{3}+\frac{\hat{S}_*}{2}\bigg)+x^{7/2} \bigg[\hat{S}_*
		\bigg(-\frac{1}{8}+\frac{g_{*2}^{\rm NLO}}{3}-\frac{5 \nu }{6}\bigg)\cr&\hspace{0.5cm}+\hat{S}
		\bigg(\frac{g_2^{\rm NLO}}{3}-\frac{53 \nu }{72}\bigg)\bigg]+x^{9/2} \bigg\{\hat{S}
		\bigg[\frac{g_{2}^{\rm N^2LO}}{6}+\frac{13
			g_2^{\rm NLO}}{12}-\frac{g_{4}^{\rm N^2LO}}{20}+\frac{g_{5}^{\rm N^2LO}}{4}-\bigg(\frac{75}{1
			6}+\frac{7 g_2^{\rm NLO}}{18}\bigg) \nu -\frac{5 \nu ^2}{16}\bigg]\cr&\hspace{0.5cm}+\hat{S}_*
		\bigg[\frac{g_{*2}^{\rm N^2LO}}{6}+\frac{13
			g_{*2}^{\rm NLO}}{12}-\frac{g_{*4}^{\rm N^2LO}}{20}+\frac{g_{*5}^{\rm N^2LO}}{4}-\frac{5}{8}+\bigg(
		-\frac{25}{6}-\frac{7 g_{*2}^{\rm NLO}}{18}\bigg) \nu -\frac{\nu
			^2}{8}\bigg]\bigg\}\cr&\hspace{0.5cm}+x^{11/2} \bigg\{\hat{S}_* \bigg[\frac{4
			g_{*2}^{\rm N^2LO}}{3}+\frac{2 g_{*2}^{\rm N^3LO}}{15}+\frac{103
			g_{*2}^{\rm NLO}}{30}-\frac{13 g_{*4}^{\rm N^2LO}}{200}+\frac{9
			g_{*5}^{\rm N^2LO}}{10}+\frac{3
			g_{*5}^{\rm N^3LO}}{20}+\frac{g_{*7}^{\rm N^3LO}}{56}-\frac{g_{*8}^{\rm N^3LO}}{40}+\frac{g_{*9}^{\rm N^3LO}}{5}\cr&\hspace{0.5cm}-\frac{993}{640}+\bigg(\frac{2933}{96}-\frac{g_{*2}^{\rm N^2LO}}{4}-\frac{16
			g_{*2}^{\rm NLO}}{15}+\frac{3 g_{*4}^{\rm N^2LO}}{40}-\frac{3
			g_{*5}^{\rm N^2LO}}{8}-\frac{41 \pi ^2}{32}\bigg) \nu
		+\bigg(\frac{3511}{360}+\frac{35 g_{*2}^{\rm NLO}}{216}\bigg) \nu ^2+\frac{887 \nu
			^3}{1296}\bigg]\cr&\hspace{0.5cm}+\hat{S} \bigg[\frac{4 g_{2}^{\rm N^2LO}}{3}+\frac{2
			g_{2}^{\rm N^3LO}}{15}+\frac{103 g_2^{\rm NLO}}{30}-\frac{13 g_{4}^{\rm N^2LO}}{200}+\frac{9
			g_{5}^{\rm N^2LO}}{10}+\frac{3
			g_{5}^{\rm N^3LO}}{20}+\frac{g_{7}^{\rm N^3LO}}{56}-\frac{g_{8}^{\rm N^3LO}}{40}+\frac{g_{9}^{\rm N^3LO}}{5}\cr&\hspace{0.5cm}+\bigg(\frac{67279}{1920}-\frac{g_{2}^{\rm N^2LO}}{4}-\frac{16
			g_2^{\rm NLO}}{15}+\frac{3 g_{4}^{\rm N^2LO}}{40}-\frac{3 g_{5}^{\rm N^2LO}}{8}-\frac{37 \pi
			^2}{16}\bigg) \nu +\bigg(\frac{9691}{960}+\frac{35 g_2^{\rm NLO}}{216}\bigg) \nu
		^2+\frac{22201 \nu ^3}{31104}\bigg]\bigg\}\,,
	\end{align}
\end{widetext}
where we explicitly see that both the orbital frequency and the angular momentum are gauge-invariant quantities in the circular limit, such that $p_\varphi(x)$ loses each dependency on the gauge parameters. On the contrary, the latter are still correctly present in the $u(x)$ relation.

The periastron advance is given by \cite{Hinderer:2013uwa}
\begin{align}
	&\dfrac{\Delta \Phi}{2\pi} = K-1, 
	\\
	& K\equiv \dfrac{\Omega_\varphi}{\Omega_r} \Bigg|_{\rm circ}
	= 
	\bigg(\dfrac{\partial^2 \hat{H}_{\rm eff}}{\partial r^2} \, \dfrac{\partial^2 \hat{H}_{\rm eff}}{\partial p_r^2}\bigg)^{-1}
	\dfrac{\partial \hat{H}_{\rm eff}}{\partial p_\varphi}
	\Bigg|_{\rm circ},
\end{align}
where the circular limit is taken as in $\hat{H}_{\rm EOB}^{\rm circ}(x)$ above, that is by taking $p_r=0$ and substituting $p_\varphi$ and $u$ with their series expansion in $x$, Eqs.~\eqref{eq:pphi} and \eqref{eq:u} respectively.

Our results for ${E}_b$ and ${K}$ read
\begin{widetext}
	\begin{align} \label{eq:Eb}
		&E_b = -\frac{x}{2}+x^2 \bigg(\frac{3}{8}+\frac{\nu }{24}\bigg)+x^3
		\bigg(\frac{27}{16}-\frac{19 \nu }{16}+\frac{\nu ^2}{48}\bigg)+x^4
		\bigg[\frac{675}{128}+\bigg(-\frac{34445}{1152}+\frac{205 \pi ^2}{192}\bigg) \nu
		+\frac{155 \nu ^2}{192}+\frac{35 \nu ^3}{10368}\bigg]
		\cr&\hspace{0.5cm}
		+x^5
		\bigg[\frac{3969}{256}+\nu 
		\bigg(\frac{123671}{11520}-\frac{448 \gamma_{\rm E} }{15}-\frac{9037 \pi ^2}{3072}-\frac{896
			\log 2}{15}-\frac{224 \log x}{15}\bigg)+\bigg(\frac{498449}{6912}-\frac{3157 \pi ^2}{1152}\bigg) \nu
		^2\cr&\hspace{0.5cm}-\frac{301 \nu ^3}{3456}-\frac{77 \nu ^4}{62208}\bigg]-x^{5/2} \bigg(\frac{4}{3} 
		\hat{S}+\hat{S}_*\bigg)+x^{7/2}
		\bigg[\bigg(-4+\frac{31 \nu }{18}\bigg)
		\hat{S}+\bigg(-\frac{3}{2}+\frac{5 \nu }{3}\bigg)
		\hat{S}_*\bigg]\cr&\hspace{0.5cm}+x^{9/2} \bigg[\bigg(-\frac{27}{2}+\frac{211
			\nu }{8}-\frac{7 \nu ^2}{12}\bigg)
		\hat{S}+\bigg(-\frac{27}{8}+\frac{39 \nu }{2}-\frac{5 \nu
			^2}{8}\bigg) \hat{S}_*\bigg]\cr&\hspace{0.5cm}+x^{11/2}
		\bigg\{\bigg[-45+\bigg(\frac{19679  }{144}+\frac{29 \pi ^2 
		}{24}\bigg)\nu-\frac{1979 \nu ^2}{36}-\frac{265 \nu ^3}{3888}\bigg]
		\hat{S}+\bigg(-\frac{135}{16}+\frac{565 \nu }{8}-\frac{1109
			\nu ^2}{24}-\frac{25 \nu ^3}{324}\bigg) \hat{S}_*\bigg\},
	\end{align}
	\begin{align}
		\label{eq:K}
		&K=1+3 x+x^2 \bigg(\frac{27}{2}-7 \nu \bigg)+x^3
		\bigg[\frac{135}{2}+\bigg(-\frac{649}{4}+\frac{123 \pi
			^2}{32}\bigg) \nu +7 \nu ^2\bigg]+x^4
		\bigg[\frac{2835}{8}+\nu\bigg(-\frac{275941}{360}-\frac{2512 \gamma_{\rm E} }{15}\cr&\hspace{0.5cm}+\frac{48007
			\pi ^2}{3072}-\frac{592 \log 2}{15}-\frac{1458 \log
			3}{5}-\frac{1256 \log x}{15}\bigg)+\bigg(\frac{5861}{12}-\frac{451 \pi
			^2}{32}\bigg) \nu ^2-\frac{98 \nu ^3}{27} 
		\bigg] \cr&\hspace{0.5cm} +x^{3/2}
		\bigg(-4 \hat{S}-3 \hat{S}_*\bigg)+x^{5/2}
		\bigg[\bigg(-34+\frac{17 \nu }{2}\bigg)
		\hat{S}+\bigg(-18+\frac{15 \nu }{2}\bigg)
		\hat{S}_*\bigg]+x^{7/2} \bigg[\bigg(-252+\frac{5317 \nu
		}{24}-\frac{22 \nu ^2}{3}\bigg)
		\hat{S}\cr&\hspace{0.5cm}+\bigg(-\frac{243}{2}+\frac{1313 \nu }{8}-7 \nu
		^2\bigg) \hat{S}_*\bigg]+x^{9/2}
		\bigg\{\bigg[-1755+\bigg(\frac{504173}{144}-\frac{3655 \pi
			^2}{96}\bigg) \nu -\frac{4419 \nu ^2}{8}+3 \nu ^3\bigg]
		\hat{S}\cr&\hspace{0.5cm}+\bigg[-810+\bigg(\frac{111401}{48}-\frac{533 \pi
			^2}{16}\bigg) \nu -\frac{3661 \nu ^2}{8}+3 \nu ^3\bigg]
		\hat{S}_*\bigg\},
	\end{align}
\end{widetext}
where we see that, in compliance with the gauge-invariance, all the 20 gauge coefficients of $\big(G_S,G_{S_{*}}\big)$ have correctly disappeared.

Furthermore, we notice that Eq.~\eqref{eq:Eb} reproduces the binding energy for circular orbits computed from the DJS gauge in Ref.~\cite{Antonelli:2020ybz}.
Instead, Eq.~\eqref{eq:K} extends to 4.5PN the orbital and spin-orbit component of the periastron advance computed in Ref.~\cite{LeTiec:2013uey}.

\section{Gauge choices}
\label{sec:N3LO_gauge_fixing}

From the general expressions of Eqs.~\eqref{eq:gauge_unfixed_GS}-\eqref{eq:gauge_unfixed_GSstar}, by suitably fixing the gauge parameters we can obtain the $\rm N^3LO$ expressions of $\big(G_S,G_{S_{*}}\big)$ in any spin gauge we want.
 For each gauge choice, we will also discuss the corresponding factorized and resummed prescriptions for $\big(G_S,G_{S_{*}}\big)$, in the spirit of Ref.~\cite{Damour:2014sva}, and compare the final results.

\subsection{The $\rm \bold{DJS}$ gauge}
\label{subsec:DJSgauge}

To recover the DJS spin gauge from our general expressions, we can simply fix the gauge coefficients so that the dependence on $p^2$ of $\big(G_S,G_{S_{*}}\big)$ is completely removed, making them functions of just $u$ and $p_r$.
This would result in the formulas computed in Ref.~\cite{Antonelli:2020aeb} when choosing \textit{a priori} the DJS gauge.
In Appendix~\ref{subapp:coeff_DJS} we collect the associated choices for the gauge coefficients, while explicit results for $\big(g_S^{\rm DJS},g_{S_{*}}^{\rm DJS}\big)$ can be found in Appendix~\ref{app:gs_PNexp}. 

Inspired by the prescription for $\big(G_S,G_{S_{*}}\big)$ proposed in Sec.~IIIB of Ref.~\cite{Damour:2014sva} (and currently  used in \TEOBResumS), we can reorganize the analytical information of the PN-expanded gyro-gravitomagnetic functions in a similarly factorized and resummed form. We can thus write
\begin{subequations}
	\begin{align}
	\label{eq:factGS_DJS}
	G_S^{\rm DJS} &= G_S^{\rm DJS,0} \hat{G}^{\rm DJS}_S, \\
	\label{eq:factGSstar_DJS}
	G_{S_{*}}^{\rm DJS} &= G_{S_{*}}^{\rm DJS,0} ~\hat{G}^{\rm DJS}_{S_{*}},
\end{align} 
\end{subequations}
where the two prefactors read
\begin{equation}
	\label{eq:djs_pref}
	G_S^{\rm DJS,0} =2 u u_c^2, \qquad G_{S_{*}}^{\rm DJS,0} = \frac{3}{2} u^3,
\end{equation} 
and correspond to the leading orders of $\big(G_S^{\rm DJS},G_{S_{*}}^{\rm DJS}\big)$. Notice that while $G_S^{\rm DJS,0}$ coincide with the spinning-particle gyro-gravitomagnetic function \eqref{eq:kerr_GS}, $G_{S_{*}}^{\rm DJS,0}$ is just the leading order in the PN expansion of Eq.~\eqref{eq:kerr_GSstar}.
For reference, once any spin-cube term is neglected by replacing $u_c$ with $u$, the latter reads
\begin{widetext}
	\begin{align}
		\label{eq:kerr_GSstar_expanded}
		G^K_{S_*} &= u^3\bigg[ 
		\frac{3}{2} - \frac{1}{c^2} \bigg(\frac{5 p^2}{8}+\frac{u}{2} \bigg) +\frac{1}{c^4} \bigg(\frac{7 p^4}{16}+\frac{p^2 u}{4}+\frac{5 p_r^2 u}{4}
		-\frac{u^2}{2}\bigg) \cr &\hspace{0.5cm}+\frac{1}{c^6}\bigg(-\frac{45 p^6}{128}-\frac{3 p^4 u}{16}-\frac{7}{4} p^2 p_r^2 u+\frac{p^2
			u^2}{4}-\frac{p_r^2 u^2}{2}-\frac{5 u^3}{8}\bigg) + \mathcal{O}\bigg(\frac{1}{c^8}\bigg)
		\bigg].
	\end{align}
\end{widetext}
The reason behind this choice for $G^0_{S_*}$ is that $G^K_{S_*}$ in the DJS spin gauge presents a singularity at the light ring location. 
This make the spin particle information in $G^K_{S_*}$  usable only in PN-expanded form, as it is done (in the circular components) in the current spin-orbit prescription of \TEOBResumS.
We also stress that the test-mass limit $\nu\to0$ of the DJS result for $G_{S_*}$ (which is provided in Appendix~\ref{app:gs_PNexp}) differs from Eq.~\eqref{eq:kerr_GSstar_expanded} at every PN order beyond the leading one, thus making the choice of shaping the prefactor $G^0_{S_*}$ after $G^K_{S_*}$ less natural.

Coming to the corresponding PN-correcting factors $\big(\hat{G}^{\rm DJS}_S,\hat{G}^{\rm DJS}_{S_*}\big)$, they are inverse-resummed (see \cite{Damour:2014sva}) as
\begin{subequations}
	\begin{align}
	&\hat{G}^{\rm DJS}_S \equiv  \dfrac{1}{T_{\rm 3PN}\left[\big(G_S^{\rm DJS}/G_S^{{\rm DJS},0}\big)^{-1}\right]},
	\\
	\label{eq:GSstar_DJS_factroized}
	&\hat{G}^{\rm DJS}_{S_*} \equiv  \dfrac{1}{T_{\rm 3PN}\left[\big(G_{S_*}^{\overline{\rm DJS}_{\rm Kerr}}/G_{S_*}^{\overline{\rm DJS}_{\rm Kerr},0}\big)^{-1}\right]},
\end{align}
\end{subequations}
where the operator $T_{\rm 3PN}$ denotes a PN Taylor expansion up to the (relative) third order.
Explicitly, we find\footnote{For simplicity, spin-cube terms are always neglected in our $\rm N^3LO$ correcting factors.}
\begin{widetext}
	\begin{subequations}
	\begin{align}
		&\big(\hat{G}^{{\rm DJS}}_{S}\big)^{-1} = 1+\nu \bigg\{\frac{1}{c^2}\bigg(
		\frac{27 p_r^2}{16}+\frac{5 u}{16}
		\bigg)+
		\dfrac{1}{c^4}\bigg[
		\bigg(\frac{649}{256}-\frac{35 \nu }{16}\bigg)
		p_r^4+\bigg(\frac{807}{128}-\frac{23 \nu }{16}\bigg) p_r^2
		u+\bigg(\frac{1657}{256}+\frac{\nu }{16}\bigg) u^2
		\bigg]\cr&\hspace{0.5cm}+
		\dfrac{1}{c^6}\bigg[
		\bigg(\frac{15251}{4096}-\frac{819 \nu }{128}+\frac{665 \nu ^2}{256}\bigg)
		p_r^6+\bigg(\frac{70215}{4096}-\frac{1229 \nu }{64}+\frac{771 \nu
			^2}{256}\bigg) p_r^4 u+\bigg(\frac{272649}{4096}-\frac{1579 \nu }{32}-\frac{69
			\nu ^2}{128}\bigg) p_r^2 u^2\cr&\hspace{0.5cm}+\bigg(\frac{1434389}{36864}-\frac{753 \nu
		}{128}+\frac{7 \nu ^2}{256}-\frac{241 \pi ^2}{384}\bigg) u^3
		\bigg] \bigg\}, \\\cr
		\label{eq:GSstar_DJS_PN_factor}
		&\big(\hat{G}^{{\rm DJS}}_{S_*}\big)^{-1}= 1+\frac{1}{c^2}\bigg[\bigg(\frac{5}{4}+\frac{3 \nu }{2}\bigg) p_r^2+\bigg(\frac{3}{4}+\frac{\nu
		}{2}\bigg) u\bigg]+
		\dfrac{1}{c^4}\bigg[
		\bigg(\frac{5}{48}+\frac{25 \nu }{12}+\frac{3 \nu ^2}{8}\bigg) p_r^4+\bigg(-1+5
		\nu -\frac{7 \nu ^2}{8}\bigg) p_r^2 u\cr&\hspace{0.5cm}+\bigg(\frac{27}{16}+\frac{29 \nu
		}{4}+\frac{3 \nu ^2}{8}\bigg) u^2
		\bigg]+
		\dfrac{1}{c^6}\bigg\{
		\bigg(-\frac{5}{96}+\frac{5 \nu }{16}+\frac{37 \nu ^2}{32}-\frac{\nu ^3}{16}\bigg)
		p_r^6+\bigg(-\frac{5}{48}+\frac{5 \nu }{2}+\frac{107 \nu ^2}{96}-\frac{9 \nu
			^3}{16}\bigg) p_r^4 u\cr&\hspace{0.5cm}+\bigg(\frac{69}{64}+\frac{1337 \nu }{32}-\frac{217 \nu
			^2}{8}-2 \nu ^3\bigg) p_r^2 u^2+\bigg[\frac{135}{32}-\frac{5 \nu
			^2}{32}+\frac{5 \nu ^3}{16}+\nu  \bigg(\frac{5501}{144}-\frac{41 \pi
			^2}{48}\bigg)\bigg] u^3
		\bigg\},
	\end{align}
	\end{subequations}
\end{widetext}
which correspond to the current \TEOBResumS{} prescription with the addition of an $\rm N^3LO$ term (proportional to $1/c^6$) linked to the results of Ref.~\cite{Antonelli:2020aeb}.

Notice that, not being able to factor out the complete spinning-particle prefactor in Eq.~\eqref{eq:GSstar_DJS_factroized}, the PN factor \eqref{eq:GSstar_DJS_PN_factor} does not reduce to 1 in the limit $\nu \to 0$.

\subsection{The $\rm \bold{\overline{DJS}}$ gauge}
\label{subsec:antiDJSguage}

In this section, we choose an alternative spin gauge (see also Ref.~\cite{Rettegno:2019tzh}),
 specifically defined around the condition that the limit $\nu \to 0$ of $G_{S_{*}}$ exactly reduces to $G_{S_{*}}^K$ and does not present the light-ring singularities that appeared in the DJS gauge.
 It turns out that imposing this condition on Eq.~\eqref{eq:gauge_unfixed_GSstar} is equivalent to the requirement that any term containing both the radial momentum and $\nu$ should disappear, and this is actually sufficient to fix all the gauge parameters also in \eqref{eq:gauge_unfixed_GS}. 
 We denote this gauge as ${\rm \overline{DJS}}$.
 The resulting PN-expanded gyro-gravitomagnetic functions $\big(g_S^{\rm \overline{DJS}},g_{S_{*}}^{\rm \overline{DJS}}\big)$ can be found in Appendix~\ref{app:gs_PNexp}, while the relative gauge coefficients are 
 collected in Appendix~\ref{subapp:coeff_antiDJS}.

We can now factorize the PN-expanded expressions as we did for the 
DJS-gauge formulas, Eqs.~\eqref{eq:factGS_DJS} and \eqref{eq:factGSstar_DJS}.
The advantage of the ${\rm \overline{DJS}}$ gauge is that we are able to factorize the full $G_{S_{*}}^K$, Eq.~\eqref{eq:kerr_GSstar}, instead of just its LO.
This means that in the test-mass limit, $\nu \to 0$, the ${\rm \overline{DJS}}$ functions will exactly reduce to their spinning-particle equivalent, while the DJS ones will only recover their PN expansion in the circular limit.

The factorization procedure is complicated by the fact that $G_{S_{*}}^0$ is no longer as simple as before but contains structures like the metric potentials $A^K$ and $D^K$.
There is hence an inherent ambiguity in extending it to comparable-mass systems,
with the only constraint that the general prefactor must reduce to $G_{S_{*}}^K$ in the probe limit $\nu \rightarrow 0$.
In order to be as agnostic as possible, we explore two different possibilities: 
(i) keeping the Kerr prefactor, with the $A^K$ and $D^K$ potentials, and introducing $\nu$-dependent terms only in the residual PN series;
(ii) promoting the metric potentials in the prefactor to the comparable-mass EOB potentials $A$ and $D$, splitting the $\nu$-dependent terms between the prefactor and the residual PN corrections.
We will denote these two different $G_{S_*}$, respectively, by $G_{S_*}^{\overline{\rm DJS}_{\rm Kerr}}$ and $G_{S_*}^{\overline{\rm DJS}_{\rm EOB}}$.

Let us start by discussing the former choice of keeping the Kerr prefactor. 
Given the $\rm N^3LO$ expressions \eqref{eq:antiDJS_GS}-\eqref{eq:antiDJS_GSstar} in $\overline{\rm DJS}$ gauge, we can define the associated prescription
\begin{subequations}
	\begin{align}
	\label{eq:factGS_antiDJS}
	G_S^{\overline{\rm DJS}} &= G_S^{\overline{\rm DJS},0}~ \hat{G}^{\overline{\rm DJS}}_S, \\
	\label{eq:factGSstar_antiDJS}
	G_{S_{*}}^{\overline{\rm DJS}_{\rm Kerr}} &= G_{S_{*}}^{\overline{\rm DJS}_{\rm Kerr},0} ~\hat{G}^{\overline{\rm DJS}_{\rm Kerr}}_{S_{*}},
\end{align}
\end{subequations}
where $G_S^{\overline{\rm DJS},0} \equiv 2 u u_c^2$ as in DJS gauge but $G_{S_{*}}^{\overline{\rm DJS}_{\rm Kerr},0}\equiv G_{S_*}^{K}(u_c^K \to u_c)$. 
The two PN-correcting factors are again inverse-resummed as
\begin{subequations}
	\begin{align}
	 &\hat{G}^{\overline{\rm DJS}}_S \equiv  \dfrac{1}{T_{\rm 3PN}\left[\big(G_S^{\overline{\rm DJS}}/G_S^{\overline{\rm DJS},0}\big)^{-1}\right]},
	 \\
	 &\hat{G}^{\overline{\rm DJS}_{\rm Kerr}}_{S_*} =  \dfrac{1}{T_{\rm 3PN}\left[\big(G_{S_*}^{\overline{\rm DJS}_{\rm Kerr}}/G_{S_*}^{\overline{\rm DJS}_{\rm Kerr},0}\big)^{-1}\right]},
\end{align}
\end{subequations}
and explicitly read
\begin{widetext}
\begin{subequations}
	\begin{align}
	&\big(\hat{G}^{\overline{\rm DJS}}_{S}\big)^{-1} = 1+\nu \bigg\{\frac{1}{c^2}\bigg(
	\frac{9  }{16}p^2-\frac{ 1 }{4}u
	\bigg)+
	\dfrac{1}{c^4}\bigg[
	 \bigg(-\frac{1 }{16}-\frac{31 \nu }{256}\bigg)p^4+ \bigg(\frac{33  }{32}+\frac{11
		\nu }{32}\bigg)p^2 u+ \bigg(\frac{119  }{32}-\frac{\nu }{16}\bigg)u^2
	\bigg]\cr&\hspace{0.5cm}+
	\dfrac{1}{c^6}\bigg[
	 \bigg(-\frac{1 }{256}+\frac{9 \nu }{128}+\frac{233 \nu ^2}{4096}\bigg)p^6+
	\bigg(-\frac{1 }{4}-\frac{31 \nu }{256}-\frac{291 \nu ^2}{1024}\bigg)p^4 u+
	\bigg(\frac{1231  }{160}-\frac{2201 \nu }{1280}+\frac{87 \nu ^2}{256}\bigg)p^2 u^2\cr&\hspace{0.5cm}+
	\bigg( \frac{28331}{1440}-\frac{241 \pi ^2}{384} +\frac{389 \nu
		}{320}-\frac{\nu ^2}{64}\bigg)u^3
	\bigg] \bigg\}, \\\cr
	&\big(\hat{G}^{\overline{\rm DJS}_{\rm Kerr}}_{S_*}\big)^{-1}= 1+\nu \bigg\{\frac{1}{c^2}\frac{p^2}{2}+
	\dfrac{1}{c^4}\bigg[
	 \bigg(-\frac{1}{8}-\frac{\nu }{8}\bigg) p^4+\bigg(\frac{11}{12}+\frac{\nu
	}{4}\bigg)p^2 u + \bigg(\frac{55}{12}+\frac{\nu }{4}\bigg)u^2
	\bigg]
	\cr&\hspace{0.5cm}+
	\dfrac{1}{c^6}\bigg[
	\bigg(\frac{1}{16}+\frac{\nu }{16}+\frac{\nu ^2}{16}\bigg)
	p^6+\bigg(\frac{1}{144}-\frac{5 \nu }{48}-\frac{\nu ^2}{4}\bigg) p^4 u-\frac{5}{12}
	p^2 p_r^2 u+\bigg(\frac{847}{144}-\frac{559 \nu }{240}\bigg) p^2
	u^2
	\cr&\hspace{0.5cm}+
	\bigg(21-\frac{41 \pi ^2}{48}+\frac{103 \nu }{30}+\frac{\nu ^2}{2}\bigg) u^3
	\bigg]\bigg\}.
\end{align}
\end{subequations}
\end{widetext}

We now consider the second option discussed above and promote the metric potentials entering $G_{S_{*}}^{K}$ to the EOB potentials, Eqs.~\eqref{eq:Aeob} and \eqref{eq:Deob}.
The procedure is exactly the same, but this time the prefactor $G_{S_{*}}^{\overline{\rm DJS}_{\rm EOB},0}$ is given by $G_{S_{*}}^K(A^K\to A, D^K \to D, u_c^K \to u_c)$.  
In this case the inverse-resummed PN correction to $G_{S_{*}}^{\overline{\rm DJS}_{\rm EOB}}$ reads
\begin{widetext}
	\begin{align} \label{eq:antiDJS_GSstar_eob}
		&\big(\hat{G}^{\overline{\rm DJS}_{\rm EOB}}_{S_*}\big)^{-1}= 1+\nu \bigg\{\frac{1}{c^2}\bigg(
		\frac{1  }{2}p^2-2 u
		\bigg)+
		\dfrac{1}{c^4}\bigg[
		\bigg(-\frac{1}{8}-\frac{\nu }{8}\bigg) p^4+\bigg(\frac{13  }{12}-\frac{3 \nu }{4}\bigg) p^2 u+\bigg(-\frac{121 }{12}+\frac{9 \nu }{4}\bigg) u^2
		\bigg]
		\cr&\hspace{0.5cm}+
		\dfrac{1}{c^6}\bigg[
		\bigg(\frac{1 }{16}+\frac{\nu }{16}+\frac{\nu ^2}{16}\bigg) p^6+\bigg(-\frac{13 }{144}+\frac{11 \nu }{48}\bigg) p^4 u-\frac{5}{12}   p^2p_r^2 u+ \bigg(\frac{1031 }{144}-\frac{933 \nu }{80}+\frac{\nu ^2}{2}\bigg) p^2u^2-\frac{17}{6}   p_r^2 u^2\cr&\hspace{0.5cm}+\bigg(\frac{4328}{135}-\frac{1184 \gamma_{\rm E} }{45}+\frac{25729 \pi ^2}{4608}+\frac{6496 \log 2}{45}-\frac{972 \log 3}{5}-\frac{592\log u}{45}+  \bigg(\frac{398}{5}-\frac{41 \pi ^2}{16}\bigg)\nu \bigg) u^3
		\bigg] \bigg\}.
	\end{align}
\end{widetext}

While both prescriptions for $G_{S_{*}}$ could be justified from the
analytical point of view, we find they both have issues.
On the one hand, we know the Kerr prefactor $G_{S_{*}}^{\overline{\rm DJS}_{\rm Kerr},0}$ could cause problems for comparable-mass systems, because the Kerr metric potentials become singular at the Kerr event horizons while, from other studies, we know that the EOB event horizons can move inwards
or even disappear when considering comparable-mass BHs.
On the other hand, the EOB PN-correction, Eq.~\eqref{eq:antiDJS_GSstar_eob}, has a more complicated structure than its corresponding Kerr correction, Eq.~\eqref{eq:antiDJS_GSstar}.
Comparing these equations we can see how $\hat{G}_{S_{*}}^{\overline{\rm DJS}_{\rm EOB}}$ contains both an additional term, proportional to $u/c^2$, and a transcendental N$^3$LO coefficient, needed to combine with 
the 4PN terms of the prefactor to give a rational PN expansion for the full $G_{S_{*}}$.
These seem to be indications that we are factoring out some structure not present in the full general relativity result.
Keeping these criticisms in mind, we now move on to discuss the effects due to these gauge choices.



\subsection{Comparing gauges}
\label{sec:testing_GS_and_GSstar}

\begin{figure}[t!]
	\includegraphics[width=0.48\textwidth]{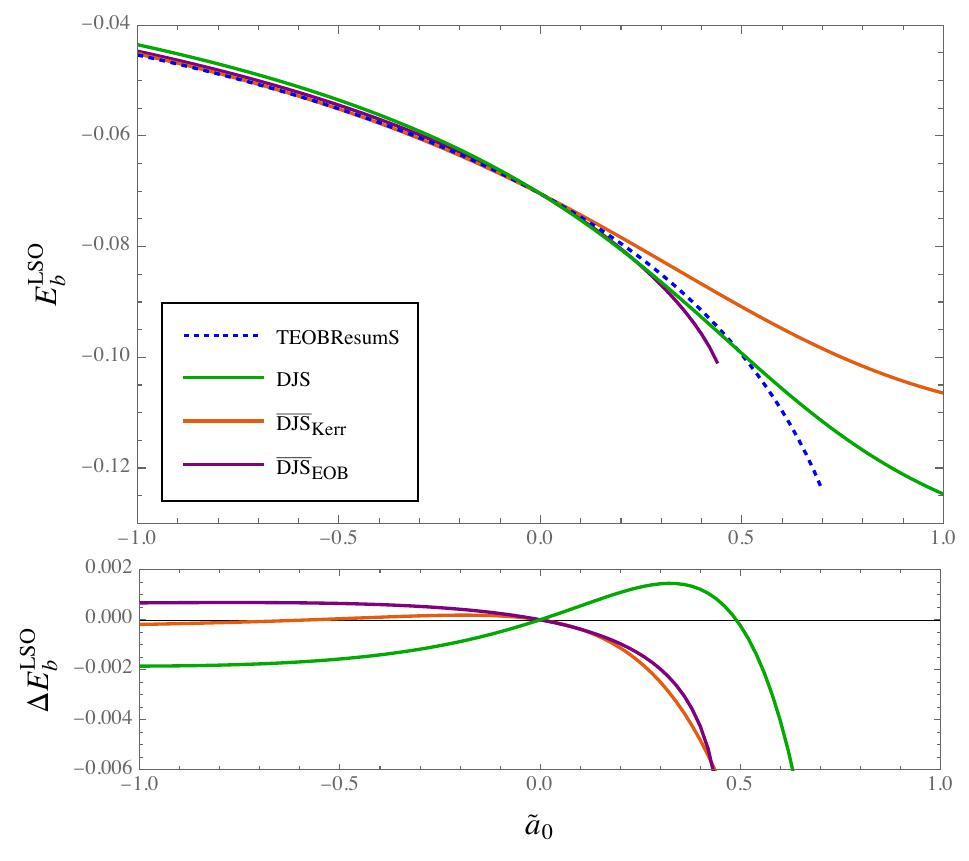}
	\caption{\label{fig:comparing_different_prescriptions} 
		Binding energy at the LSO versus $\tilde{a}_0$ in the equal-mass, equal-spin case. 
		Top panel: comparison between the binding energy of the EOB model \TEOBResumS{} with its NR-calibrated spin-orbit sector (blue dashed line) to different $\rm N^3LO$-accurate prescriptions of the gyro-gravitomagnetic functions:
		(i) DJS gauge (green line);
		(ii) Kerr-factorized $\rm \overline{DJS}$ gauge functions (red line);
		and (iii) EOB-factorized $\rm \overline{DJS}$ gauge (purple line).
		Each curve stops at the value of $\tilde{a}_0$ for which the LSO no longer exists. 
		Bottom panel: binding energy difference between the various $\rm N^3LO$ curves and the corresponding NR-calibrated \TEOBResumS{} value.}
\end{figure}

\begin{figure}[t!]
	\includegraphics[width=0.48\textwidth]{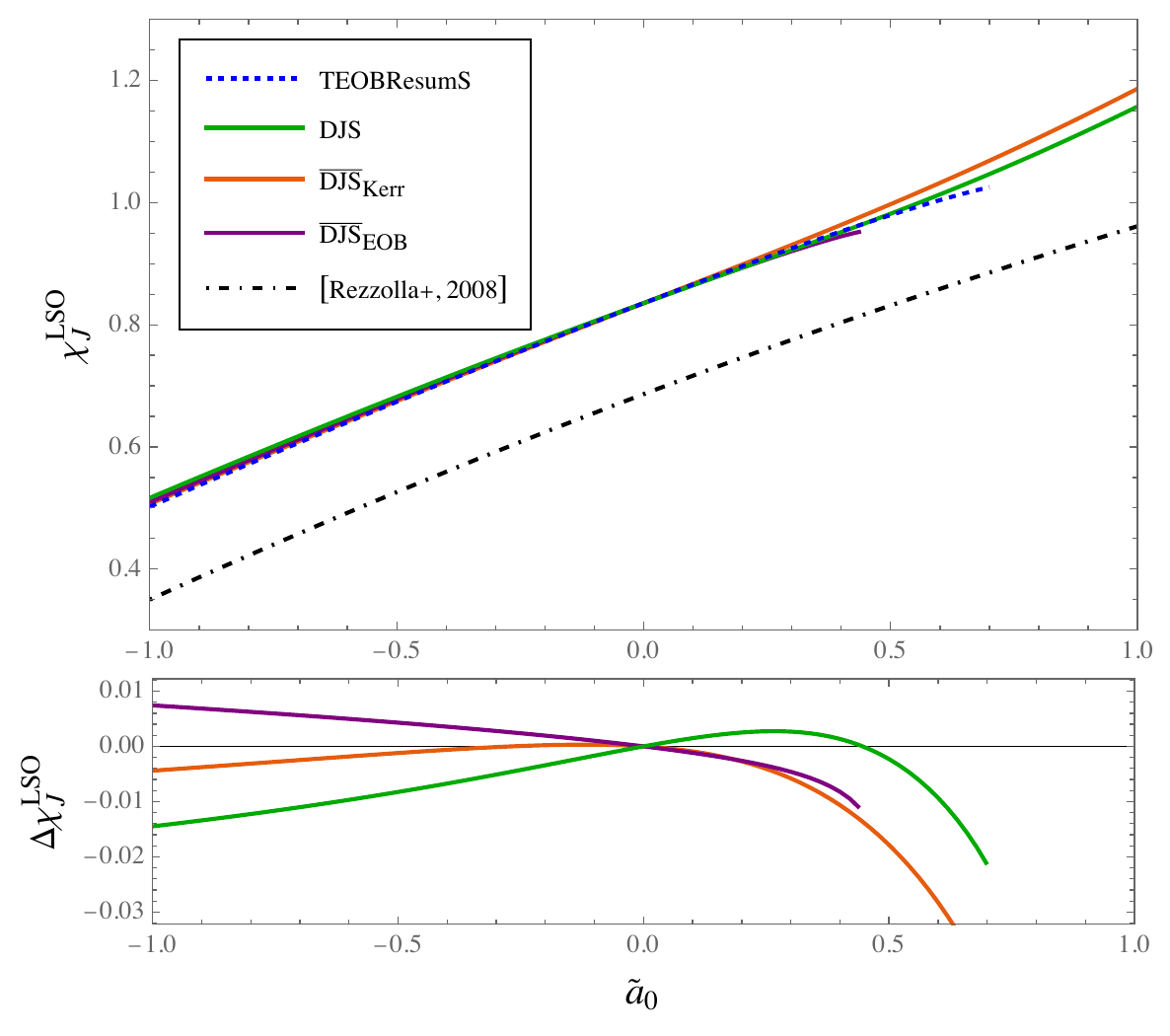}
	\caption{\label{fig:chiJ_LSO__rc_NLO} 
		Analogue of Fig.~\ref{fig:comparing_different_prescriptions} for the Kerr parameter $\chi_J$ at the LSO. 
		We also add the the numerical result of Ref.~\cite{Rezzolla:2007xa} (dash-dotted black line), which represents the Kerr parameter of the residual BH after the coalescence.
		The loss of angular momentum between the LSO and merger 
		is responsible for the lower values of this quantity. 
	}
\end{figure}

Here we study the behavior of gauge-invariant quantities at the last stable orbit (LSO), which are informative about the general characteristics of the dynamics. 

We will focus on the equal-mass ($\nu=1/4$) and equal-spin ($\hat{S}=\hat{S_*}=\tilde{a}_0/2$) case and study the variation, changing $\tilde{a}_0$, of the LSO values of the binding energy $E_b$, Eq.~\eqref{eq:Eb_def}, and the dimensionless Kerr parameter
\begin{equation} \label{eq:kerr_parameter}
	\chi_J \equiv \frac{1}{\nu} \frac{j_{\rm tot}}{\hat{H}_{\rm EOB}^{\rm circ}}\,,
\end{equation}
where the total angular momentum $j_{\rm tot}$ is given (in the equal-mass, equal-spin case) by
\begin{equation}
	j_{\rm tot} = p_\varphi + 2 \tilde{a}_0.
\end{equation}
Here, we do not PN-expand these quantities, as we did in Eq.~\eqref{eq:Eb}, so that the results will depend on the gauge choices we make.

We recall that the LSO radius $r_{\rm LSO}$ can be found by solving numerically the system of equations
\begin{equation}
	\frac{\partial \hat{H}_{\rm eff}}{\partial r}\Big|_{r=r_{\rm LSO}} =0, \qquad \frac{\partial^2 \hat{H}_{\rm eff}}{\partial r^2}\Big|_{r=r_{\rm LSO}}=0.
\end{equation}

Fig.~\ref{fig:comparing_different_prescriptions} compares the binding energy at the LSO for the current version of \TEOBResumS~\cite{Nagar:2023zxh} and three alternative $\rm N^3LO$-accurate spin-orbit prescriptions, introduced above.
We will use here the NR-informed \TEOBResumS{} as a good approximation
of quantities which are not easy to extract from numerical simulations themselves.

We recall that \TEOBResumS{} contains NR-informed parameters both in the orbital and the spin-orbit sectors (see Ref.~\cite{Nagar:2023zxh} and references therein for more details).
For the analytical N$^3$LO spin-orbit contributions, we keep the same 
(NR-calibrated) orbital effective Hamiltonian and change only 
$G_S$ and $G_{S_*}$, so to isolate their effect.
This implies that each curve will overlap when $\tilde{a}_0 \rightarrow 0$.

Notably, without any numerical calibration in the spin-orbit part, the $\rm \overline{DJS}$ curves are very close (at least for negative spins) to the dynamics calibrated to numerical relativity simulations.
This is somewhat surprising, given that the \TEOBResumS{} gyro-gravitomagnetic functions are expressed in DJS gauge. 
These are in fact equivalent
to the N$^3$LO DJS functions computed here up to 3.5PN and differ at the 4.5PN level, where an NR-fitted parameter was used in place of the unknown (at the time) analytical coefficient\footnote{There is a subtlety. In \TEOBResumS{}~\cite{Nagar:2023zxh,Rettegno:2019tzh}, the DJS series are expressed as functions of $(p_{r_*},u_c$) instead of $(p_r,u)$.}.
The fact that our DJS curve lies further apart from the \TEOBResumS{} one implies that the analytical N$^3$LO coefficient [see Eq.~\eqref{eq:GSstar_DJS_PN_factor}] has a very different effect with respect to the NR-calibrated one, which only contains a term proportional to $u^3/c^6$~\cite{Nagar:2023zxh}.

We also note that the $\rm \overline{DJS}_{\rm Kerr}$ and DJS curves predict a LSO for each value of the spin variable, while the \TEOBResumS{} LSO and $\rm \overline{DJS}_{\rm EOB}$ ones stop existing for $\tilde{a}_0 \gtrsim 0.7$ and $\tilde{a}_0 \gtrsim 0.45$ respectively. 

We show in Fig.~\ref{fig:chiJ_LSO__rc_NLO} the behavior of the Kerr parameter $\chi_J$ at the LSO for varying $\tilde{a}_0$.
Again, we can observe how the previous DJS prescription used in \TEOBResumS~is quite close to both the $\rm N^3LO$-accurate $\rm \overline{DJS}$ prescriptions of Sec.~\ref{subsec:antiDJSguage}. 
In Fig.~\ref{fig:chiJ_LSO__rc_NLO} we also show for reference the numerically simulated Kerr parameter \emph{after} the coalescence, as obtained from the analytic fit of Ref.~\cite{Rezzolla:2007xa}. The fact that the latter is systematically below our results for $\chi_J^{\rm LSO}$ is in agreement with the loss of angular momentum during the plunge and merger.

These results suggest that the inclusion of the newly computed terms, in $\rm \overline{DJS}$ gauge, could improve the accuracy of EOB-NR models for spinning binaries.

\section{Conclusions}
\label{sec:conclusions}

In this paper we have computed the $G_S$ and $G_{S_*}$ functions,
which determine the spin-orbit interaction of a BBH system, at 4.5PN-level in gauge-unfixed form. We hence extended the results of Ref.~\cite{Antonelli:2020aeb}, which performed these computations by specifying \textit{ab initio} the DJS spin gauge. 

By comparing analytical scattering-angle information to the corresponding quantity computed using a parameterized EOB Hamiltonian, we were able to 
extract the gauge-unfixed form of the gyro-gravitomagnetic functions.
We then specified these both in DJS gauge (as in Ref.~\cite{Antonelli:2020aeb}), defined so as to make $G_S$ and $G_{S_*}$ independent of the angular momentum, and in the alternative $\rm \overline{DJS}$ gauge.
This spin-gauge is defined requiring that in the test-mass limit ($\nu\to0$) $G_{S_*}$ reduces exactly to $G^K_{S_*}$ [Eq.~\eqref{eq:kerr_GSstar}], the function describing the spin-orbit interaction between a spinning particle and a Kerr BH.
This is impossible in DJS gauge because of coordinate singularities at 
the light-ring location. 

We used this computations to extend the PN knowledge of the periastron advance for quasi-circular-orbits binaries by one perturbative order, that is up to the 4.5PN [see Eq.~\eqref{eq:K}].

Based on the procedure followed by the \TEOBResumS{} model, we then proposed and tested a factorized prescription for $\big(G_S,G_{S_{*}}\big)$ in each gauge.
We factorize the LO contribution and inverse-resum the residual PN correction, so to tame its behavior in the strong-field regime.
While this is straightforward for the DJS gauge, where the LO of each gyro-gravitomagnetic function is simple [see Eq.~\eqref{eq:djs_pref}],
it becomes more difficult when discussing the $\rm \overline{DJS}$ gauge.
In the latter gauge, the LO prefactor $G^K_{S_*}$ contains the Kerr
metric potentials which have a direct extension to comparable masses in the EOB framework.
We can then choose to: (i) factorize $G^K_{S_*}$, with the original Kerr potentials, which we dubbed Kerr-factorized $\overline{\rm DJS}$ gauge; or (ii)
promote the Kerr functions in $G^K_{S_*}$ to the full comparable-mass EOB equivalents, determining the EOB-factorized $\overline{\rm DJS}$ gauge.
Both these choices have drawbacks.
The Kerr-factorized functions impose coordinate singularities at the location of Kerr event horizons, although we know 
that the EOB horizon location depends on the mass ratio.
Conversely, in the EOB-factorized $\overline{\rm DJS}$ gauge, the residual 
PN series contains transcendental terms, needed to compensate the ones present in the EOB metric potentials at 4PN level.

We compared in Figs.~\ref{fig:comparing_different_prescriptions} and \ref{fig:chiJ_LSO__rc_NLO} two LSO quantities for each gauge choice against the NR-informed spin-orbit sector of the \TEOBResumS{} model, that is known to have a very good agreement with NR simulations of compact binaries. 
We find that quantities computed in each $\overline{\rm DJS}$ spin-orbit prescription lie very close to the NR-informed \TEOBResumS{} ones.

We think these comparisons indicate that the newly-computed terms, when expressed in a suitable gauge, could help improve the overall accuracy of
EOB-based models for spinning compact binaries in their inspiral phase.
We only considered here the conservative dynamics, which is directly modified by the addition of 4.5PN spin-orbit terms.
A deeper study is needed, considering also GW fluxes and waveforms and the eventual inclusion of the (incomplete) 5.5PN spin-orbit information provided in Ref.~\cite{Khalil:2021fpm}. 
Finally, it is also worth testing whether the analytical N$^3$LO $\overline{\rm DJS}$-gauge functions computed here can be suitably NR-informed using an extra N$^4$LO effective parameter, so to further improve the description of the spin-orbit interactions in the strong-field regime.

\section*{Acknowledgments}
The authors thank the hospitality and the stimulating environment of the Institut des Hautes Etudes Scientifiques. 
P.~R. is supported by the Italian Minister of University and Research (MUR) via the 
PRIN 2020KB33TP, {\it Multimessenger astronomy in the Einstein Telescope Era (METE)}.
The present research was also partly supported by the ``\textit{2021 Balzan Prize for 
	Gravitation: Physical and Astrophysical Aspects}'', awarded to Thibault Damour.

\appendix
\section{Coefficients of the gyro-gravitomagnetic functions
}
In this Appendix we provide explicitly the  conditions we find on the coefficients of $G^{\rm gen}_{S}$ and $G^{\rm gen}_{S_*}$ (see Eq.~\eqref{eq:generalGS}) once we impose the matching with the 3PN scattering angle and after we fix the spin gauge according to the DJS and $\overline{\rm DJS}$ choices.
\subsection{Relations from the scattering angle matching}
\label{app:coeff_rel_matching}
Here we collect the coefficient relations that are obtained by enforcing the matching between the spin orbit parts of the effective and 3PN scattering angles, discussed in Sec.~\ref{subsec:gauge_unfixed_GS_and_GSstar} of the main text. They are
\begin{widetext}
	\begin{subequations}
	\begin{align}
		&g_{1}^{\rm NLO}= -\frac{g_{2}^{\rm NLO}}{3}-\frac{9 \nu }{8}, \qquad g_{3}^{\rm NLO}= \frac{g_{2}^{\rm NLO}}{3}+\frac{\nu }{2}, \\\cr
			&
			g_{1}^{\rm N^2LO}= -\frac{g_{2}^{\rm N^2LO}}{3}-\frac{g_{4}^{\rm N^2LO}}{5}+\frac{7 \nu ^2}{8}+\frac{\nu }{8} ,
			\\& 
			g_{3}^{\rm N^2LO}= \frac{g_{2}^{\rm N^2LO}}{2}+\frac{g_{2}^{\rm NLO}}{4}+\frac{9 g_{4}^{\rm N^2LO}}{20}-\frac{g_{5}^{\rm N^2LO}}{4}-\frac{5 \nu ^2}{4}-\frac{33 \nu }{16},
			\\&
			g_{6}^{\rm N^2LO}= -\frac{g_{2}^{\rm N^2LO}}{6}+\frac{3 g_{2}^{\rm NLO}}{4}-\frac{g_{4}^{\rm N^2LO}}{4}+\frac{g_{5}^{\rm N^2LO}}{4}+\frac{\nu ^2}{4}-\frac{119 \nu }{16}, \\\cr
			&
			g_{1}^{\rm N^3LO}= -\frac{g_{2}^{\rm N^3LO}}{3}-\frac{g_{4}^{\rm N^3LO}}{5}-\frac{g_{7}^{\rm N^3LO}}{7}-\frac{95 \nu ^3}{128}-\frac{9 \nu ^2}{32}+\frac{\nu }{128},
			\\& 
			g_{3}^{\rm N^3LO}= \frac{g_{2}^{\rm N^2LO}}{3}+\frac{2 g_{2}^{\rm N^3LO}}{3}+\frac{3 g_{4}^{\rm N^2LO}}{8}+\frac{3 g_{4}^{\rm N^3LO}}{5}-\frac{g_{5}^{\rm N^3LO}}{4}+\frac{29 g_{7}^{\rm N^3LO}}{56}-\frac{g_{8}^{\rm N^3LO}}{8}+\frac{31 \nu ^3}{16}+\frac{21 \nu ^2}{8}+\frac{\nu }{2},
			\\& 
			g_{6}^{\rm N^3LO}= \frac{2 g_{2}^{\rm N^2LO}}{3}-\frac{7 g_{2}^{\rm N^3LO}}{15}+\frac{2 g_{2}^{\rm NLO} \nu }{5}+\frac{17 g_{2}^{\rm NLO}}{30}+\frac{63 g_{4}^{\rm N^2LO}}{200}-\frac{3 g_{4}^{\rm N^3LO}}{5}+\frac{g_{5}^{\rm N^2LO}}{10}+\frac{7 g_{5}^{\rm N^3LO}}{20}-\frac{5 g_{7}^{\rm N^3LO}}{8}\cr&\hspace{0.5cm}+\frac{11 g_{8}^{\rm N^3LO}}{40}-\frac{g_{9}^{\rm N^3LO}}{5}-\frac{11 \nu ^3}{8}+\frac{431 \nu ^2}{40}-\frac{1231 \nu }{80},
			\\& 
			g_{10}^{\rm N^3LO}= -\frac{g_{2}^{\rm N^2LO}}{2}+\frac{2 g_{2}^{\rm N^3LO}}{15}-\frac{7 g_{2}^{\rm NLO} \nu }{5}+\frac{101 g_{2}^{\rm NLO}}{60}-\frac{21 g_{4}^{\rm N^2LO}}{25}+\frac{g_{4}^{\rm N^3LO}}{5}+\frac{13 g_{5}^{\rm N^2LO}}{20}-\frac{g_{5}^{\rm N^3LO}}{10}+\frac{g_{7}^{\rm N^3LO}}{4}\cr&\hspace{0.5cm}-\frac{3 g_{8}^{\rm N^3LO}}{20}+\frac{g_{9}^{\rm N^3LO}}{5}+\frac{\nu ^3}{8}-\frac{123 \nu ^2}{20}+\frac{241 \pi ^2 \nu }{192}-\frac{28331 \nu }{720},
	\end{align}
\end{subequations}
and 
	\begin{subequations}
	\begin{align}
		&g_{*1}^{\rm NLO} = -\frac{g_{*2}^{\rm NLO}}{3}-\frac{3 \nu }{4}-\frac{5}{8}, \qquad g_{*3}^{\rm NLO} = \frac{g_{*2}^{\rm NLO}}{3}-\frac{1}{2}, \\\cr
			&g_{*1}^{\rm N^2LO}= -\frac{g_{*2}^{\rm N^2LO}}{3}-\frac{g_{*4}^{\rm N^2LO}}{5}+\frac{9 \nu ^2}{16}+\frac{\nu }{2}+\frac{7}{16}, \\& 
			g_{*3}^{\rm N^2LO}= \frac{g_{*2}^{\rm N^2LO}}{2}+\frac{g_{*2}^{\rm NLO}}{4}+\frac{9 g_{*4}^{\rm N^2LO}}{20}-\frac{g_{*5}^{\rm N^2LO}}{4}-\frac{3 \nu ^2}{8}-\frac{9 \nu }{8}+\frac{9}{16}, \\&
			g_{*6}^{\rm N^2LO}= -\frac{g_{*2}^{\rm N^2LO}}{6}+\frac{3 g_{*2}^{\rm NLO}}{4}-\frac{g_{*4}^{\rm N^2LO}}{4}+\frac{g_{*5}^{\rm N^2LO}}{4}-\frac{3 \nu ^2}{8}-\frac{55 \nu }{8}-\frac{13}{16}, \\\cr
			&
			g_{*1}^{\rm N^3LO}= -\frac{g_{*2}^{\rm N^3LO}}{3}-\frac{g_{*4}^{\rm N^3LO}}{5}-\frac{g_{*7}^{\rm N^3LO}}{7}-\frac{15 \nu ^3}{32}-\frac{33 \nu ^2}{64}-\frac{25 \nu }{64}-\frac{45}{128},
			\\&
			g_{*3}^{\rm N^3LO}= \frac{g_{*2}^{\rm N^2LO}}{3}+\frac{2 g_{*2}^{\rm N^3LO}}{3}+\frac{3 g_{*4}^{\rm N^2LO}}{8}+\frac{3 g_{*4}^{\rm N^3LO}}{5}-\frac{g_{*5}^{\rm N^3LO}}{4}+\frac{29 g_{*7}^{\rm N^3LO}}{56}-\frac{g_{*8}^{\rm N^3LO}}{8}+\frac{3 \nu ^3}{4}+\frac{3 \nu ^2}{2}+\frac{3 \nu }{8}-\frac{5}{8},
			\\&
			g_{*6}^{\rm N^3LO}= \frac{2 g_{*2}^{\rm N^2LO}}{3}-\frac{7 g_{*2}^{\rm N^3LO}}{15}+\frac{2 g_{*2}^{\rm NLO} \nu }{5}+\frac{17 g_{*2}^{\rm NLO}}{30}+\frac{63 g_{*4}^{\rm N^2LO}}{200}-\frac{3 g_{*4}^{\rm N^3LO}}{5}+\frac{g_{*5}^{\rm N^2LO}}{10}+\frac{7 g_{*5}^{\rm N^3LO}}{20}-\frac{5 g_{*7}^{\rm N^3LO}}{8}\cr &\hspace{0.5cm}+\frac{11 g_{*8}^{\rm N^3LO}}{40}-\frac{g_{*9}^{\rm N^3LO}}{5}+\frac{3 \nu ^3}{8}+\frac{213 \nu ^2}{20}-\frac{21 \nu }{4}+\frac{51}{80},
			\\&
			g_{*10}^{\rm N^3LO}= -\frac{g_{*2}^{\rm N^2LO}}{2}+\frac{2 g_{*2}^{\rm N^3LO}}{15}-\frac{7 g_{*2}^{\rm NLO} \nu }{5}+\frac{101 g_{*2}^{\rm NLO}}{60}-\frac{21 g_{*4}^{\rm N^2LO}}{25}+\frac{g_{*4}^{\rm N^3LO}}{5}+\frac{13 g_{*5}^{\rm N^2LO}}{20}-\frac{g_{*5}^{\rm N^3LO}}{10}+\frac{g_{*7}^{\rm N^3LO}}{4}\cr &\hspace{0.5cm}-\frac{3 g_{*8}^{\rm N^3LO}}{20}+\frac{g_{*9}^{\rm N^3LO}}{5}-\frac{3 \nu ^3}{4}-\frac{201 \nu ^2}{40}+\frac{41 \pi ^2 \nu }{32}-\frac{701 \nu }{24}-\frac{121}{80}.
\end{align}
\end{subequations}
\end{widetext}
\subsection{Coefficient choices for the $\rm DJS$ gauge}
\label{subapp:coeff_DJS}

From the general gauge-unfixed expressions for $\big(G_S,G_{S_{*}}\big)$ given in Eqs.~\eqref{eq:gauge_unfixed_GS}-\eqref{eq:gauge_unfixed_GSstar} we can determine the corresponding expressions in any well-defined  spin gauge we want, by suitably fixing the gauge coefficients. 

For the DJS expressions \eqref{eq:DJS_GS}-\eqref{eq:DJS_GSstar} the associated coefficient choices are
\begin{subequations}
\begin{align}
&g_2^{\rm NLO}= -\frac{27 \nu }{8}, \\\cr
\begin{split}
&g_2^{\rm N^2LO}= 0, \quad
g_4^{\rm N^2LO}= \frac{5 \nu }{8}+\frac{35 \nu ^2}{8}, \cr
&g_5^{\rm N^2LO}= -\frac{21 \nu }{2}+\frac{23 \nu ^2}{8}, 
\end{split}
\\\cr
\begin{split}
&g_2^{\rm N^3LO}= 0, \quad 
g_4^{\rm N^3LO}= 0, \quad
g_5^{\rm N^3LO}= 0, \cr
&g_7^{\rm N^3LO}= \frac{7 \nu }{128}-\frac{63 \nu ^2}{32}-\frac{665 \nu ^3}{128}, \cr&
g_8^{\rm N^3LO}= \frac{781 \nu }{128}+\frac{831 \nu ^2}{32}-\frac{771 \nu ^3}{128}, \cr&
g_9^{\rm N^3LO}= -\frac{5283 \nu }{64}+\frac{1557 \nu ^2}{16}+\frac{69 \nu ^3}{64},
\end{split}
\end{align}
\end{subequations}
and 
\begin{subequations}
	\begin{align}
	\begin{split}
&g_{*2}^{\rm NLO}= -\frac{15}{8}-\frac{9 \nu }{4},
	\end{split}
\\\cr
	\begin{split}
	&g_{*2}^{\rm N^2LO}= 0, \quad
	g_{*4}^{\rm N^2LO}= \frac{35}{16}+\frac{5 \nu }{2}+\frac{45 \nu ^2}{16}, \cr&
	g_{*5}^{\rm N^2LO}= \frac{69}{16}-\frac{9 \nu }{4}+\frac{57 \nu ^2}{16},
	\end{split}
      \\\cr
	\begin{split}
		&g_{*2}^{\rm N^3LO}= 0,
		\quad
		g_{*4}^{\rm N^3LO}= 0,
		\quad
		g_{*5}^{\rm N^3LO}= 0,
		\cr&
		g_{*7}^{\rm N^3LO}= -\frac{315}{128}-\frac{175 \nu }{64}-\frac{231 \nu ^2}{64}-\frac{105 \nu ^3}{32},
		\cr&
		g_{*8}^{\rm N^3LO}= -\frac{1105}{128}-\frac{53 \nu }{64}+\frac{351 \nu ^2}{64}-\frac{243 \nu ^3}{32},
		\cr&
		g_{*9}^{\rm N^3LO}= -\frac{45}{64}-\frac{837 \nu }{32}+\frac{2361 \nu ^2}{32}+\frac{27 \nu ^3}{16}.
	\end{split}
	\end{align}
\end{subequations}

\subsection{Coefficient choices for the $\rm \overline{DJS}$ gauge}
\label{subapp:coeff_antiDJS}
The coefficient conditions associated to the $\overline{\rm DJS}$ expressions \eqref{eq:antiDJS_GS}-\eqref{eq:antiDJS_GSstar} are much simpler, with all the gauge coefficient of $G_S$ set to zero and
\begin{subequations}
	\begin{align}
		&g_{*2}^{\rm NLO}= 0,\\\cr
		\begin{split}
			&g_{*2}^{\rm N^2LO}= 0,
			\quad
			g_{*4}^{\rm N^2LO}= 0,
			\quad
			g_{*5}^{\rm N^2LO}= \frac{5}{4},
		\end{split}\\\cr
	\begin{split}
		&g_{*2}^{\rm N^3LO}= 0,
		\quad
		g_{*4}^{\rm N^3LO}= 0,
		\quad
		g_{*5}^{\rm N^3LO}= -\frac{7}{4},
		\cr&
		g_{*7}^{\rm N^3LO}= 0,
		\quad
		g_{*8}^{\rm N^3LO}= 0,
		\quad
		g_{*9}^{\rm N^3LO}= -\frac{1}{2}.
	\end{split}
	\end{align}
\end{subequations}

\section{PN-expanded results for the gyro-gravitomagnetic functions}
\label{app:gs_PNexp}

In this appendix we provide explicitly the PN results we find for $(G_S,G_{S_*})$ after their general expressions \eqref{eq:gauge_unfixed_GS}-\eqref{eq:gauge_unfixed_GSstar} are specified to the DJS and the $\rm \overline{DJS}$ spin gauge. We adopt the same notation as in Eqs.~\eqref{eq:gauge_unfixed_GS}-\eqref{eq:gauge_unfixed_GSstar}.

In the DJS gauge our result reproduce what was found in Ref.~\cite{Antonelli:2020ybz} and reads
\begin{widetext}
	\begin{subequations}
		\begin{align} \label{eq:DJS_GS}
			&g_S^{\rm DJS} = 2+\nu\bigg\{- \dfrac{1}{c^2}\bigg( 
			\frac{27   }{8}p_r^2+\frac{5  }{8} u
			\bigg)+
			\dfrac{1}{c^4}\bigg[
			\bigg(\frac{5  }{16}+\frac{35 \nu  }{16}\bigg)
			p_r^4+\bigg(-\frac{21   }{4}+\frac{23 \nu  }{16}\bigg)
			p_r^2 u+\bigg(-\frac{51   }{8}-\frac{\nu  }{16}\bigg) u^2
			\bigg] \cr&\hspace{0.5cm}+
			\dfrac{1}{c^6}\bigg[
			\bigg(\frac{7   }{128}-\frac{63 \nu  }{32}-\frac{665 \nu
				^2}{128}\bigg) p_r^6+\bigg(\frac{781   }{128}+\frac{831
				\nu  }{32}-\frac{771 \nu ^2}{128}\bigg) p_r^4
			u+\bigg(-\frac{5283   }{64}+\frac{1557 \nu  }{16}+\frac{69
				\nu ^2}{64}\bigg) p_r^2 u^2\cr&\hspace{0.5cm}+\bigg(
			-\frac{80399}{1152}+\frac{241 \pi ^2}{192}
			+\frac{379 \nu}{32}-\frac{7 \nu ^2}{128}\bigg) u^3
			\bigg]\bigg\},\\\cr
			\label{eq:DJS_GSstar}
			&g_{S_*}^{\rm DJS} =\frac{3}{2}- \dfrac{1}{c^2}\bigg[
			\bigg(\frac{15}{8}+\frac{9 \nu }{4}\bigg)
			p_r^2+\bigg(\frac{9}{8}+\frac{3 \nu }{4}\bigg) u
			\bigg]+
			\dfrac{1}{c^4}\bigg[
			\bigg(\frac{35}{16}+\frac{5 \nu }{2}+\frac{45 \nu ^2}{16}\bigg)
			p_r^4+\bigg(\frac{69}{16}-\frac{9 \nu }{4}+\frac{57 \nu
				^2}{16}\bigg) p_r^2 u \cr&\hspace{0.5cm}+\bigg(-\frac{27}{16}-\frac{39 \nu
			}{4}-\frac{3 \nu ^2}{16}\bigg) u^2
			\bigg]+
			\dfrac{1}{c^6}\bigg\{
			\bigg(-\frac{315}{128}-\frac{175 \nu }{64}-\frac{231 \nu
				^2}{64}-\frac{105 \nu ^3}{32}\bigg)
			p_r^6 \cr&\hspace{0.5cm}+\bigg(-\frac{1105}{128}-\frac{53 \nu }{64}+\frac{351
				\nu ^2}{64}-\frac{243 \nu ^3}{32}\bigg) p_r^4
			u+\bigg(-\frac{45}{64}-\frac{837 \nu }{32}+\frac{2361 \nu
				^2}{32}+\frac{27 \nu ^3}{16}\bigg) p_r^2
			u^2 \cr&\hspace{0.5cm}+\bigg[-\frac{405}{128}+\bigg(-\frac{7627}{192}+\frac{41
				\pi ^2}{32}\bigg) \nu +\frac{711 \nu ^2}{64}-\frac{3 \nu
				^3}{32}\bigg] u^3
			\bigg\}.
		\end{align}
	\end{subequations}
\end{widetext}

Their equivalent in the $\rm \overline{DJS}$ gauge is instead given by
\begin{widetext}
	\begin{subequations}
		\begin{align} \label{eq:antiDJS_GS}
			&g_{S}^{\rm \overline{DJS}} = 2+ \nu \bigg\{\dfrac{1}{c^2}\bigg(
			-\frac{9  }{8} p^2+\frac{u}{2}
			\bigg)+
			\dfrac{1}{c^4}\bigg[
			\bigg(\frac{1 }{8}+\frac{7 \nu}{8}\bigg) p^4+\bigg(-\frac{33  }{16}-\frac{5 \nu }{4}\bigg) p^2 u+\bigg(-\frac{119 }{16}+\frac{\nu}{4}\bigg) u^2
			\bigg]\cr&\hspace{0.5cm}+
			\dfrac{1}{c^6}\bigg[
			\bigg(\frac{1 }{128}-\frac{9 \nu}{32}-\frac{95 \nu ^2}{128}\bigg) p^6+\bigg(\frac{1 }{2}+\frac{21 \nu }{8}+\frac{31 \nu ^2}{16}\bigg) p^4 u+\bigg(-\frac{1231 }{80}+\frac{431 \nu }{40}-\frac{11 \nu ^2}{8}\bigg) p^2 u^2\cr&\hspace{0.5cm}+\bigg(-\frac{28331  }{720}+\frac{241   \pi ^2}{192}-\frac{123 \nu}{20}+\frac{\nu ^2}{8}\bigg) u^3
			\bigg]\bigg\}, \\\cr
			\label{eq:antiDJS_GSstar}
			&g_{S_*}^{\rm \overline{DJS}}= \frac{3}{2}- \dfrac{1}{c^2}\bigg[
			\bigg(\frac{5}{8}+\frac{3 \nu }{4}\bigg) p^2+\frac{u}{2}
			\bigg]+
			\dfrac{1}{c^4}\bigg[
			\bigg(\frac{7}{16}+\frac{\nu }{2}+\frac{9 \nu ^2}{16}\bigg) p^4+\frac{5 p_r^2 u}{4}+\bigg(\frac{1}{4}-\frac{9 \nu }{8}-\frac{3 \nu ^2}{8}\bigg) p^2 u\cr&\hspace{0.5cm}+\bigg(-\frac{1}{2}-\frac{55 \nu }{8}-\frac{3 \nu ^2}{8}\bigg) u^2
			\bigg]+
			\dfrac{1}{c^6}\bigg\{
			\bigg(-\frac{45}{128}-\frac{25 \nu }{64}-\frac{33 \nu ^2}{64}-\frac{15 \nu ^3}{32}\bigg) p^6+\bigg(-\frac{3}{16}+\frac{3 \nu }{8}+\frac{3 \nu ^2}{2}+\frac{3 \nu ^3}{4}\bigg) p^4 u\cr&\hspace{0.5cm}-\frac{7}{4} p^2 p_r^2 u+\bigg(\frac{1}{4}-\frac{21 \nu }{4}+\frac{213 \nu ^2}{20}+\frac{3 \nu ^3}{8}\bigg) p^2 u^2-\frac{p_r^2 u^2}{2}-\bigg[\frac{5}{8}+  \bigg(\frac{701}{24}-\frac{41 \pi ^2}{32}\bigg)\nu+\frac{201 \nu ^2}{40}+\frac{3 \nu ^3}{4}\bigg] u^3
			\bigg\}.
		\end{align}
	\end{subequations}
\end{widetext}

\bibliography{refs.bib,local.bib}
\end{document}